\documentclass[
reprint,
prc,
showpacs,
 amsmath,amssymb,
 aps,
]{revtex4-1}

\usepackage{graphicx}
\usepackage{dcolumn}
\usepackage{bm}

\usepackage{xcolor}

\begin{document}

\preprint{APS/123-QED}

\title{ Photoneutron reaction cross section measurements on $^{94}$Mo and $^{90}$Zr relevant to the $\textit{p}$-process nucleosynthesis}


\author{A. Banu}
\email{Corresponding author: banula@jmu.edu}
\author{E. G. Meekins}
\altaffiliation[Present address:] {\indent Geisinger Medical Center, Danville, Pennsylvania 17822, USA}
\affiliation{Department of Physics and Astronomy, James Madison University, Harrisonburg, Virginia 22807, USA}

\author{J. A. Silano}
\altaffiliation[Present address:] {\indent Nuclear and Chemical Sciences Division, Lawrence Livermore National Laboratory, Livermore, California 94550, USA}
\author{H. J. Karwowski}
\affiliation{Triangle Universities Nuclear Laboratory, Durham, North Carolina 27708, USA}
\affiliation{University of North Carolina at Chapel Hill, Chapel Hill, North Carolina 27516, USA}

\author{S. Goriely}
\affiliation{Institut d'Astronomie et d'Astrophysique, Universit\'{e} Libre de Bruxelles, Campus de la Plaine, CP-226, 1050 Brussels, Belgium}

\date{\today}

\begin{abstract}

       The photodisintegration cross sections for the $^{94}$Mo($\gamma$,n) and $^{90}$Zr($\gamma$,n) reactions have been experimentally investigated with quasi-monochromatic photon beams at the High Intensity $\gamma$-ray Source (HI$\gamma$S) facility of the Triangle Universities Nuclear Laboratory (TUNL). The energy dependence of the photoneutron reaction cross sections was measured with high precision from the respective neutron emission thresholds up to 13.5 MeV. These measurements contribute to a broader investigation of nuclear reactions relevant to the understanding of the $p$-process nucleosynthesis. The results are compared with the predictions of Hauser-Feshbach statistical model calculations using two different models for the dipole $\gamma$-ray strength function. The resulting $^{94}$Mo($\gamma$,n) and $^{90}$Zr($\gamma$,n) photoneutron stellar reaction rates as a function of temperature in the typical range of interest for the $p$-process nucleosynthesis show how sensitive the photoneutron stellar reaction rate  can be to the experimental data in the vicinity of the neutron threshold.

\end{abstract}

\maketitle

\section{\label{sec:level1}Introduction}

\indent How the nuclear reactions that occur in stars and in stellar explosions have been forging the elements out of hydrogen and helium leftover from the Big Bang is a longstanding \cite{Burb57}, still timely research topic in nuclear astrophysics. Although there is a fairly complete understanding of the production of elements up to iron by nuclear fusion reactions in stars, important details concerning the production of the elements beyond iron remain puzzling. Current understanding is that the nucleosynthesis beyond iron proceeds mainly via neutron capture reactions and subsequent $\beta^{-}$ decays in the $s$- and $r$-processes. However, some 35 proton-rich stable isotopes, between $^{74}$Se and $^{196}$Hg, cannot be synthesized by neutron-capture processes since they are located on the neutron-deficient side of the valley of $\beta$-stability. They are thus shielded from the $s$- or $r$-process.\newline
\indent These proton-rich stable nuclides are generally referred to as $p$-nuclei \cite{Arno76,Woos78,Lamb92,Arno03}. As a group they are the rarest of all stable isotopes. The mechanism responsible for their synthesis is termed the $p$-process. The gross similarities between the abundance curves of the $p$-nuclei and the $s$- and $r$-nuclei imply that the much more abundant $s$- and $r$-nuclei may serve as seeds for the $p$-process, but the astrophysical details of the $p$-process are still under discussion. So far it has been impossible to reproduce the solar abundances of all $p$-isotopes using a single nucleosynthesis process. Several different sites and (independently operating) processes seem to be required, with the largest fraction of the $p$-isotopes being synthesized by sequences of photodisintegrations and $\beta^{+}$ decays. Due to the dominance of photodisintegrations, this mechanism of the $p$-process is sometime referred to as the $\gamma$-process \cite{Woos78}.\newline
\indent It is generally accepted that the $\gamma$-process occurs mainly in explosive O/Ne burning during supernova Type II explosions at temperatures in the range of T $\approx$ 2-3 GK, but supernovae Type Ia and Ib/c are also expected to contribute \cite{Arno03}. Calculations based on the $\gamma$-process concept can reproduce the bulk of the $p$-nuclei within a factor of $\approx$ 3 \cite{Raye95,Raus00}, but the most abundant $p$-isotopes, $^{92,94}$Mo and $^{96,98}$Ru (as well as potentially the A $<$ 124 mass region), are underproduced, making their nucleosynthesis one of the great outstanding mysteries in nuclear astrophysics. It is not yet clear whether specific environments need to be invoked for their production, such as He-accreting sub-Chandrasekhar white-dwarf \cite{Gori02} or $p$-rich neutrino driven winds of type-II supernovae \cite{Froh06} or $s$-process-enriched type Ia supernova \cite{Arno03,Trav11}, or if the calculated underproductions are due to deficiencies in the astrophysical models or in the underlying nuclear physics input, $i.e.$ the reaction rates used in the model.\newline 
\indent Contrary to the $s$- or $r$-process, the concepts of steady flows or reaction rate equilibria cannot be applied to the $p$-process, which operates far from equilibrium. As a result, an extended network of some 20000 reactions linking about 2000 nuclei in the A $\leq$ 210 mass region must be computed in detail \cite{Arno03}. It is impossible to measure all these reaction rates in the laboratory. Hence, it becomes obvious that the vast majority of the reaction rates must be determined theoretically. Usually the unknown reaction rates are calculated within the framework of Hauser-Feshbach (HF) statistical model calculations with typical uncertainties of about 30\% for stable nuclei \cite{Arno03,Raus00}, but that can reach a factor of 10 for neutron-deficient nuclei \cite{Arno03}. The model requires input based on nuclear structure, optical model potentials, and nuclear level densities to calculate transmission coefficients (average widths) which, in turn, determine the reaction cross sections, and thus, the reaction rates. The uncertainties involved in any HF cross section calculation are not related to the model of formation and de-excitation of the compound nucleus itself, but rather to the evaluation of the nuclear quantities necessary for the calculation of the transmission coefficients. The photon transmission coefficient is particularly relevant in the case of photonuclear reactions and is calculated assuming the dominance of dipole $E1$ transitions. The transmission coefficient for $\gamma$-ray emission with multipolarity $L$ is related to the (downward) $\gamma$-ray strength function ($\gamma$SF) $f$ as follows:
 \begin{equation}
 T_{\gamma}^{L}=2\pi E_{\gamma}^{2L+1} f(E_{\gamma}).
\end{equation}	
\indent  Much effort has been and still is devoted to measuring and understanding the electric dipole strength function that exhibits a pronounced peak at the giant dipole resonance (GDR) energy. There are many approaches used to derive $f$, each leading to an energy dependence of the $E1$ transmission coefficients given near the GDR energy by a Lorentzian function. Experimental photoabsorption data confirm the simple semi-classical prediction of a Lorentzian shape at energies near the resonance energy but this description is less satisfactory at lower energies, and especially near the photodisintegration reaction threshold \cite{Arno03}. Therefore, it is of substantial interest to develop microscopic models which are expected to provide reasonable reliability and predictive power for the $E1$-$\gamma$SF. Efforts in this direction, such as QRPA calculations \cite{Gori02b,Tson15,Mart16,Gori16,Gori18}, have been applied successfully to several photoneutron cross section measurements carried out recently with quasi-monochromatic laser-Compton scattering $\gamma$-rays \cite{Utsu09,Tonc10,Utsu13,Raut13,Fili14,Saue14,Nyhu15,Rens18,Utsu18}. \newline
\indent Despite the endemic problem of reproducing the solar abundances of $^{92,94}$Mo and $^{96,98}$Ru, as well as the part of the A $<$ 124 region, recent studies performed by Travaglio $et$ $al.$ \cite{Trav11,Trav15} in supernova Type Ia calculations using both deflagration and delayed detonation models demonstrated that both light and heavy $p$-nuclei, including the much debated isotopes $^{92}$Mo and $^{96,98}$Ru, are produced with similar enhancement factors relative to solar abundances, provided an $s$-process enrichment of the progenitor is assumed. The model, however, predicts the production of $^{94}$Mo with a much lower abundance in comparison to all the other light $p$-nuclei. Another remarkable finding of Ref. \cite{Trav11} points out that the $\gamma$-process can make important contributions to the production of the neutron magic nucleus  $^{90}$Zr, previously known as a genuine $s$-process nuclide.\newline 
\indent In light of the intriguing findings of Refs. \cite{Trav11,Trav15}, we were motivated to investigate the photoneutron reactions on $^{94}$Mo and $^{90}$Zr. The measurements were focused on studying the energy dependence of the photoneutron reaction cross sections near the respective neutron emission thresholds and up to 13.5 MeV, taking into account the fact that the energy window of effective stellar burning for photoneutron reactions is located close to the reaction threshold at $E_{\gamma}^{eff}$= ($l$+1/2)$kT $+ $S_{n}$, where $l$ is the neutron orbital angular momentum, $k$ is the universal Boltzmann constant, $T$ is the stellar temperature, and $S_{n}$ is the neutron separation energy.\newline
\indent The experimental photroneutron cross sections are compared to the predictions of HF statistical model calculations using different models for the $\gamma$SF, thus allowing the $\gamma$SF to be constrained and further to estimate the corresponding photoneutron stellar reaction rates which directly influence the $p$-process nucleosynthesis.
       
\section{Experimental details}

\indent In this experiment excitation functions of ($\gamma$,n) photodisintegration reactions on the nuclei $^{94}$Mo and $^{90}$Zr were measured close to and above the corresponding neutron emission thresholds - $S_{n}$ = 9.68 MeV for $^{94}$Mo($\gamma$,n) and $S_{n}$ = 11.97 MeV for $^{90}$Zr($\gamma$,n). The measurements reported in this paper extend to a $\gamma$-ray energy of 13.5 MeV and were performed using TUNL's High Intensity $\gamma$-ray Source (HI$\gamma$S) facility. Quasi-monoenergetic, circularly-polarized and highly intense beams of real photons, of selectable energy, were produced via intracavity back-scattering of free-electron laser (FEL) photons from relativistic electrons \cite{Well09}.\newline
\indent These beams were collimated to a diameter of 1.5 cm by a 101.6-cm long Al collimator, which was located 53 m downstream from the collision point inside the optical cavity of the HI$\gamma$S FEL storage ring and 5 m upstream from the experimental setup in the Upstream Target Room (UTR) at HI$\gamma$S. An Al collimator was used, instead of the Pb collimator generally in use at HI$\gamma$S, to limit the beam-induced neutron background in the $^{3}$He proportional counters used for the neutron detection. That is because the $^{27}$Al($\gamma$,n) reaction has a neutron emission threshold of 13.1 MeV, higher than nearly all of the $\gamma$-ray beam energies of interest in this experiment and higher than the neutron emission threshold for Pb.\newline
\indent A schematic drawing (not to scale) of the experimental setup as it was assembled for the present experiment is shown in Fig. 1.

\begin{figure*}[!]
\includegraphics[scale=0.5,keepaspectratio=true]{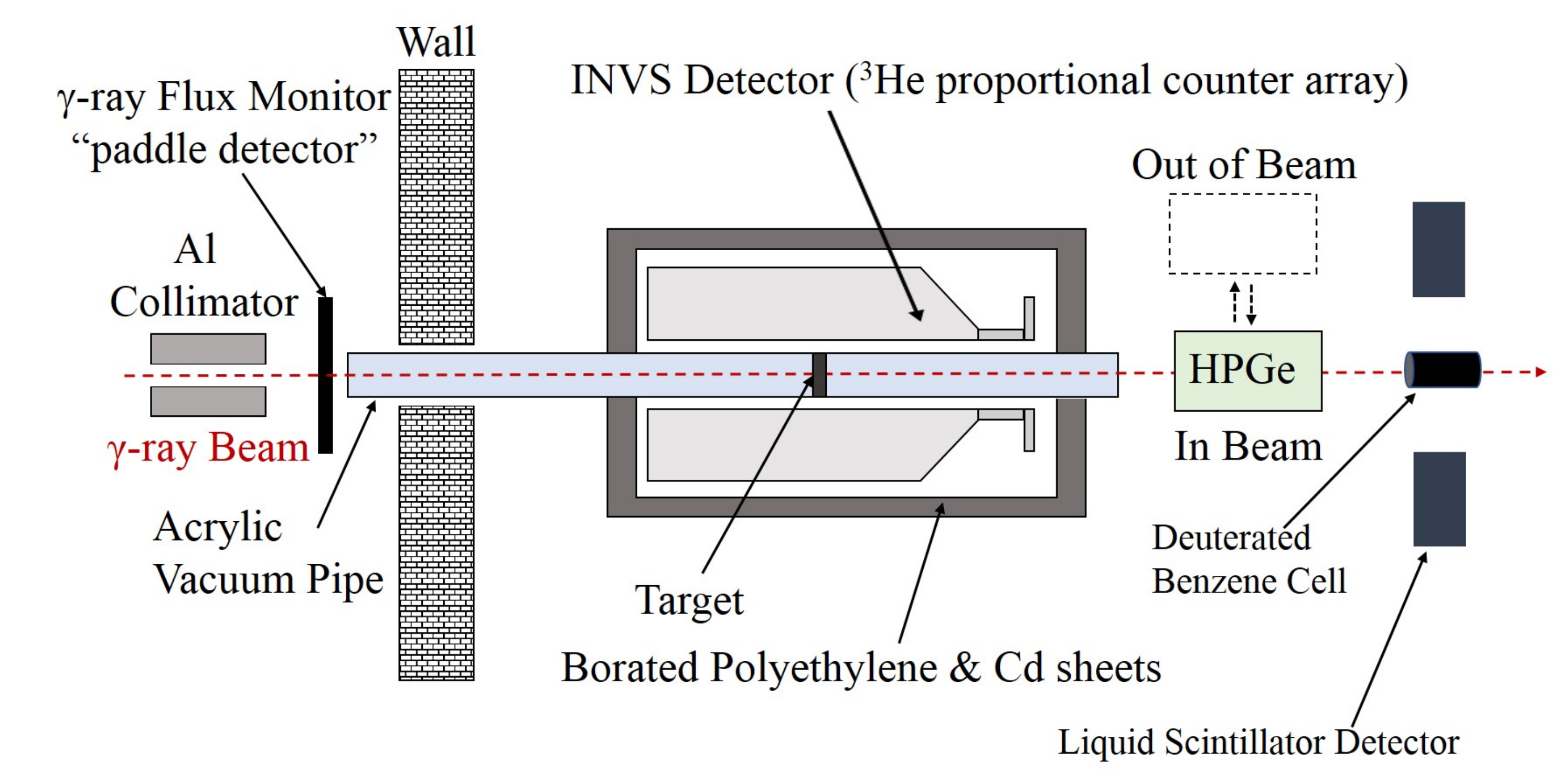}
\caption[Fig. 1]{(Color online) Schematic drawing (not to scale) of the experimental setup in the HI$\gamma$S Upstream Target Room (UTR). See the text for details of the detectors sketched in the figure.}
\end{figure*}  

The quasi-monoenergetic $\gamma$-ray beam had an energy width in the range of 4\% - 5\% (FWHM). The $\gamma$-ray beam flux was continuously monitored and yielded values in the range of 10$^{7}$- 10$^{8}$ $\gamma$/s on target. The very high $\gamma$-ray flux available at HI$\gamma$S makes this facility ideal for the investigation of photoneutron reaction cross sections with $p$-nuclei as targets.\newline
\indent The two targets of interest consisted of 98.97\% enriched $^{94}$Mo with an areal density of 598 mg/cm$^{2}$ and 97.70\% enriched $^{90}$Zr with an areal density of 1087 mg/cm$^{2}$. As illustrated in Fig. 1, the targets were mounted in the longitudinal and axial center of the INVS detector, inside a polycarbonate vacuum pipe kept under rough vacuum to prevent background due to scattering in the air of the $\gamma$-ray beam.\newline
\indent Since the $^{94}$Mo($\gamma$,n)$^{93}$Mo reaction produces the unstable residual isotope of $^{93}$Mo, which has a long half-life of $T_{1/2}$ = 3500 yr and decays by electron capture without $\gamma$-ray emission, the only way to experimentally study at HI$\gamma$S the excitation function for the $^{94}$Mo($\gamma$,n)$^{93}$Mo reaction was by direct neutron counting.\newline
\indent  The overall 98-hour beam time of the experiment was divided to accommodate beam energy measurements, ($\gamma$,n) reaction cross section measurements, as well as beam-induced background measurements of relevance for accurate neutron counting. In the following subsections the experimental details of these measurements are provided.

\subsection{\label{sec:photonbeam-energy}$\gamma$-ray beam energy measurements}

\indent The $\gamma$-ray beam energies were measured with a large high-purity germanium (HPGe) detector of 123\% relative efficiency appropriately positioned downstream of the target and the neutron detector array. As indicated in Fig. 1, the HPGe detector was remotely positioned on the beam axis for the $\gamma$-ray beam energy measurements and out of the beam axis during the ($\gamma$,n) reaction measurements. Sets of copper attenuators of precisely known thicknesses, stationed behind the exit mirror of the FEL optical cavity \cite{Well09}, were remotely inserted in the beam to reduce the $\gamma$-ray beam flux during the beam energy measurements. \newline
\indent At $\gamma$-ray beam energies above 9 MeV, of interest for this work, the measured $\gamma$-ray beam energy spectrum becomes strongly convolved with the detector response function resulting in a broad energy peak with overlapping full-energy peak, first- and second-escape peaks, and their respective Compton edges. To unfold the photon beam full-energy peak, a 7.6-cm thick segmented NaI(Tl) annulus was mounted around the 123\% HPGe detector that enabled the extraction of the $\gamma$-ray beam energy distribution. The detection of photons escaping the HPGe detector due to pair production or Compton scattering in any of the four NaI(Tl) segments of the annulus detector was recorded in anticoincidence with the HPGe detector. As shown in Fig. 2, at a $\gamma$-ray beam energy of 9 MeV, the $\gamma$-ray beam energy distribution was extracted by fitting the high-energy tail of the full-energy peak with a Gaussian function. The fit parameters which minimize the $\chi^{2}$-value of the fit give the full-energy peak of the $\gamma$-ray beam and its energy width.\newline
\indent The 123\% HPGe detector had an energy resolution of approximately 4 keV and was calibrated using the $\gamma$-ray lines from thermal neutron capture on a $^{58}$Ni target as well as from standard calibration sources and the activity of naturally occurring radioactive nuclei present in the UTR. 
\begin{figure}
\includegraphics[scale=0.45,keepaspectratio=true]{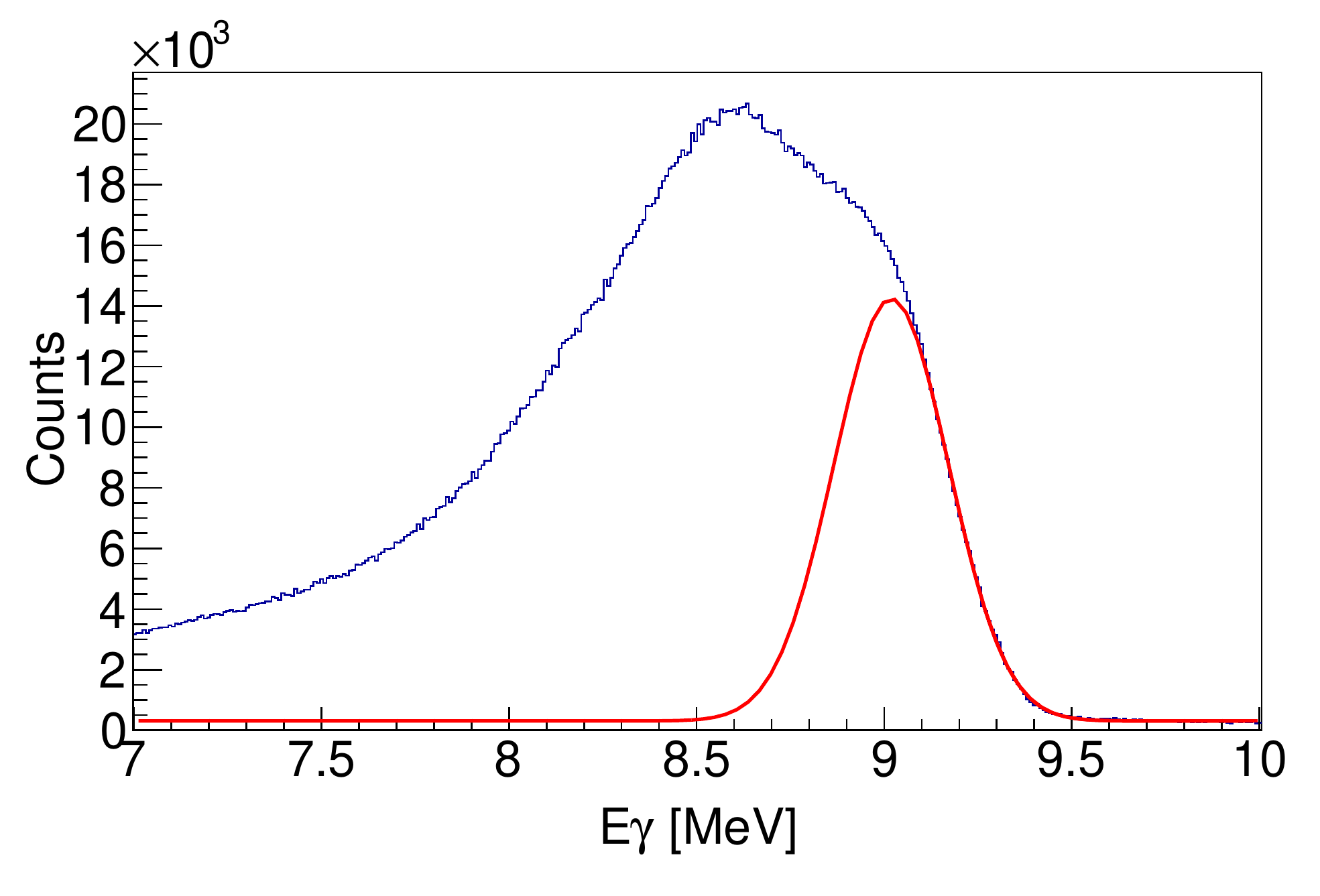}
\caption[Fig. 2]{(Color online) $\gamma$-ray spectrum recorded by the 123\% HPGe detector in anticoincidence with the NaI(Tl) annulus at a $\gamma$-ray beam energy of 9 MeV. The red curve represents a Gaussian fit from where the centroid of the full-energy peak of the $\gamma$-ray beam was determined. The peak at 8.5 MeV corresponds to the first-escape peak.}
\end{figure}  

\subsection{\label{sec:photonbeam-intensity}$\gamma$-ray beam flux measurements}

\indent The $\gamma$-ray beam flux was measured with a single Bicron BC-400 thin plastic scintillation detector (dubbed ``paddle detector" because of its geometrical shape) coupled to a photomultiplier tube \cite{Pywe09}. The paddle detector was located behind the Al collimator (see Fig. 1). By design, the scintillating paddle detects recoil electrons and positrons from the photoelectric effect, Compton and pair-production processes. Its efficiency has been shown \cite{Pywe09} to be well described by GEANT4 \cite{Gean03} simulations. \newline
\indent The flux stability in the paddle detector was monitored throughout the experiment by cross-checking it against flux measurements of the d($\gamma$,n)p reaction, which has a very well-studied  excitation function \cite{Bire85,Bern86,Scia05}. The corresponding experimental setup for the detection of neutrons from the deuteron photodisintegration consisted of two Bicron BC-501A organic liquid scintillator detectors coupled to photomultiplier tubes that were placed 46 cm from a deuterated benzene cell at a scattering angle of 90°. As illustrated in Fig. 1, the setup was located downstream at the end of the UTR with the deuterated benzene located on the axis of the $\gamma$-ray beam. 

\subsection{\label{sec:bkgrd}Neutron detection}

\indent In this experiment the neutrons were detected using an assembly of 18 tubular proportional counters, filled with $^{3}$He gas at $\sim$6 atm \cite{Arno11}. The neutron detector array, known as the model IV inventory sample counter (INVS), was originally developed at Los Alamos National Laboratory and it is presently available at TUNL. The tubes are arranged in two concentric rings of radii 7.24 cm and 10.60 cm, each containing nine equally spaced counters. The counters are embedded in a cylindrical polyethylene body 46.2 cm long and 30.5 cm in diameter which serves as a neutron moderator. A schematic drawing of the INVS detector is presented in Fig. 3.
\begin{figure}[h]
\includegraphics[scale=0.6,keepaspectratio=true]{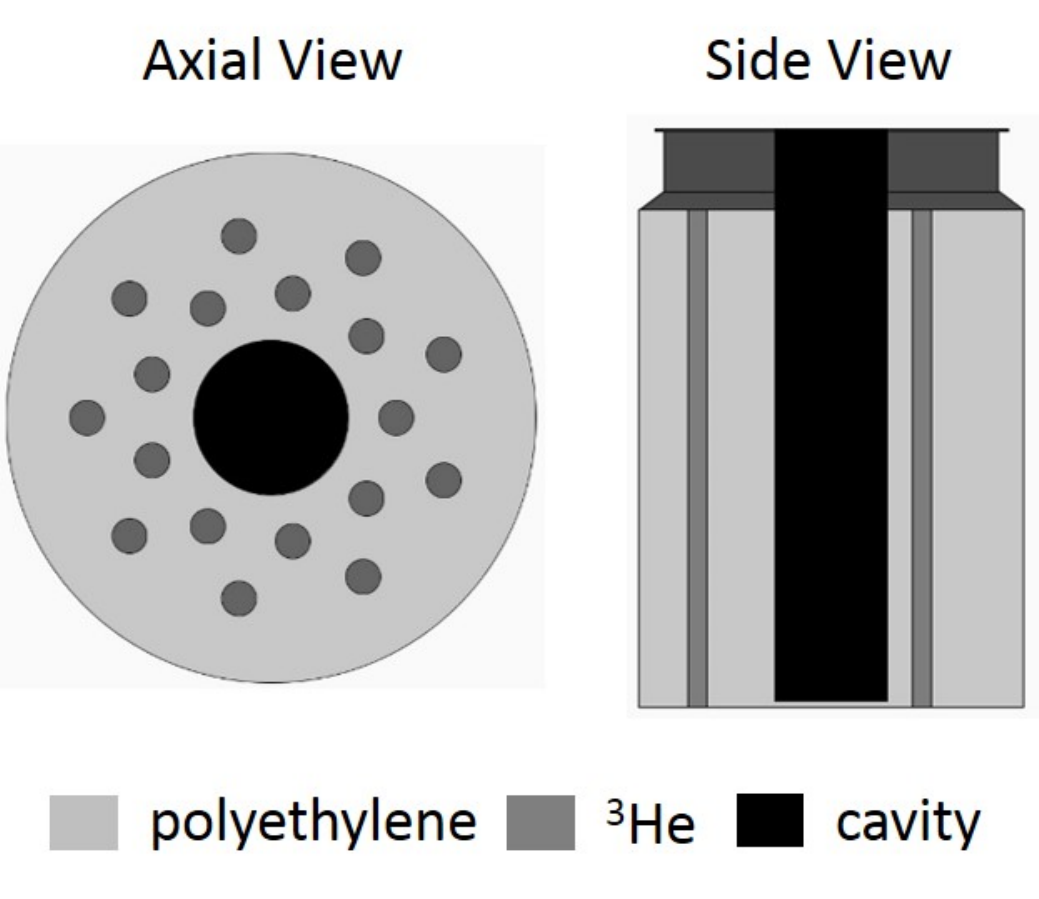}
\caption[Fig. 3]{(Color online) Axial and side cut-away views of the INVS detector showing the $^{3}$He tubes arranged in two concentric rings around a central cavity. See the text for information on dimensions. }
\end{figure}   

\indent The neutron detector array was readout by three TTL logic pulses corresponding to detections occurring in the inner ring (I), the outer ring of the array (O), and the logical OR of the I and O pulses. All three logic output signals were recorded in scalers that were integrated in the CODA ($\textbf{C}$EBAF $\textbf{O}$nline $\textbf{D}$ata $\textbf{A}$cquisition) data acquisition system. \newline
\indent To limit the rate of detection of background neutrons generated outside of the INVS detector, layers of borated polyethylene and Cd sheets were placed around it, as illustrated in Fig. 1.\newline
\indent The neutron detection efficiency depends on the neutron energy. Neutron energies of this work span a broad range from 20 keV to $\sim$4 MeV that correspond to detection efficiencies as high as $\sim$55\% and as low as $\sim$25\%, respectively. The efficiencies were simulated in GEANT4 assuming isotropically distributed neutrons, and were consistent with the experimentally measured efficiencies of Arnold $et$ $al.$ \cite{Arno11}. However, it should be noted that for the energy range studied in this work, the $^{94}$Mo($\gamma$,n) and $^{90}$Zr($\gamma$,n) reactions only proceed directly to the ground state of their respective residual nuclei at low photon beam energies. At higher energies, population of the ground state proceeds predominantly via the population of excited states in the residual nuclei which then $\gamma$ decay to the ground state. Because the information on the neutron energy is lost by the thermalization of the neutrons in the moderator, determining the neutron detection efficiency for such a detector is a complex problem to tackle. \newline 
\indent Thus, instead of determining the efficiency for the ($\gamma,n_{0}$) channel only that corresponds to neutrons emitted when the ground state in populated directly, contributions from channels which populate excited states in the daughter nucleus must also be taken into account. An effective efficiency has been defined as
\begin{equation}
\epsilon_{n}^{eff}= \sum_{i} b_{i} \epsilon_{n_{i}}(E_{n_{i}}),
 \end{equation} 
where the $b_{i}$ are the neutron branchings of the ($\gamma,n_{i}$) channel at a given $E_{\gamma}$, and $\epsilon_{n_{i}}$ are the energy-dependent detection efficiencies for neutrons from the ($\gamma,n_{i}$) channel. We calculated the neutron branching ratios $b_{i}$ of the ($\gamma,n_{i}$) channels using the TALYS nuclear reaction code \cite{Koni12} with the $\gamma$SF axially-symmetric-deformed Hartree-Fock-Bogoliubov (HFB) plus QRPA model based on the D1M Gogny interaction \cite{Mart16,Gori16,Peru08,Gori18}, the HFB plus combinatorial nuclear level density model \cite{Gori08}, and with the spherical neutron-nucleus optical-model potential of Koning and Delaroche \cite{Koni03}. More details about this theoretical framework are given in Section IV.\newline
\indent For the photon energy range reached in this experiment and under consideration of quantum mechanical selection rules we considered several contributions of the excited states in the residual $^{89}$Zr and $^{93}$Mo nuclei which can be populated by neutron $s$- and$/$or $p$-waves from $E1$ excitations of $1^{-}$ states in the $^{90}$Zr and $^{94}$Mo nuclei of interest, as presented in the following. Note: TALYS database uses the RIPL-3 library \cite{Capo09} for the treatment of the discrete levels.

\subsubsection{$^{90}$Zr($\gamma$,n)} 
\indent The highest photon beam energy reached in this experiment at $E_{\gamma}$ = 13.5 MeV will give access to an excitation window in $^{89}$Zr up to $\sim$1.5 MeV from the neutron emission threshold at $S_{n}$ = 11.97 MeV. At photon beam energies $E_{\gamma}$ $<$ $S_{n}$ + 588 keV, the $^{90}$Zr($\gamma$,n)$^{89}$Zr reaction can proceed only to the ground state of $^{89}$Zr. This path would be strongly hampered due to the large angular momentum required for the emitted neutron ($f$ wave) from the compound nucleus $^{90}$Zr$^*$ with $J^{\pi}$ = 1$^{-}$ to the ground state of $^{89}$Zr with $J^{\pi}$ = 9/2$^{+}$. Starting at photon beam energies larger than 12.8 MeV the population of the ground state in $^{90}$Zr proceeds predominantly via emitted neutrons ($s$ wave) from the first two excited states of $^{89}$Zr at 588 keV and 1.095 MeV. Table I presents the energies of the emitted neutrons with their corresponding detection efficiencies and branching ratios as simulated in GEANT4 and calculated in TALYS, respectively. The value of the effective neutron efficiency calculated from Eq. (2) for each of the photon beam energy reached in this experiment is presented in the last column of Table I.
\setlength{\tabcolsep}{0.4cm}
\begin{table*} [!]
\begin{center}
\caption{\textbf{$^{90}$Zr($\gamma$,n)$^{89}$Zr:} Photon beam energies ($E_{\gamma}$), energy levels in $^{89}$Zr ($E_{i}$), energies of emitted neutrons calculated as $E_{n_{i}} = (\frac{89}{90})(E_{\gamma} - S_{n} -E_{i}$), orbital angular momentum of the emitted neutrons ($l_{i}$), percent neutron detection efficiency at each neutron energy ($\epsilon_{n_{i}}$), neutron branching ratio for each neutron energy ($b_{i}$), and the percent effective neutron detection efficiency, as calculated from Eq. (2), for each photon beam energy ($\epsilon_{n}^{eff}$).}
\begin{tabular}{cccccccc} \hline\hline
$E_{\gamma} $ [MeV] & $E_{i} $ [MeV] &  $J_{i}^{\pi_{i}}$ &  $E_{n_{i}}$ [MeV] &  $l_{i}$ & $\epsilon_{n_{i}}$ [\%] & $b_{i}$ & $\epsilon_{n}^{eff}$ [\%] \\
\hline
 12    &  0            & 9/2$^{+}$  &  0.03  & 3 ($f$ wave) &  52.89 &  1      &  \textbf{52.89}\\
 12.1 &  0            & 9/2$^{+}$  &  0.13  & 3 ($f$ wave) &  52.15 &  1      &  \textbf{52.15}\\
 12.2 &  0            & 9/2$^{+}$  &  0.23  & 3 ($f$ wave) &  51.53 &  1      &  \textbf{51.53}\\
 12.4 &  0            & 9/2$^{+}$  &  0.43  & 3 ($f$ wave) &  49.21 &  1      &  \textbf{49.21}\\ 
 12.5 &  0            & 9/2$^{+}$  &  0.53  & 3 ($f$ wave) &  47.69 &  1      &  \textbf{47.69}\\
         &                &                  &           &                   &       &          & \\
 12.8 &  0            & 9/2$^{+}$  &  0.82  & 3 ($f$ wave) &  44.18 &  0.17 &  \textbf{49.94}\\
         &  0.5878   & 1/2$^{-}$   &  0.24  & 0 ($s$ wave) &  51.12 &  0.83 & \\
         &                &                  &           &                   &       &          & \\             
  13   &  0            & 9/2$^{+}$  &  1.02  & 3 ($f$ wave) &  41.33 &  0.23 &  \textbf{46.94}\\ 
         &  0.5878   & 1/2$^{-}$   &  0.44  & 0 ($s$ wave) &  48.61 &  0.77 & \\
         &                &                  &           &                   &       &          & \\
 13.5 &  0            & 9/2$^{+}$  &  1.51  & 3 ($f$ wave) &  36.71 &  0.26 &  \textbf{42.97}\\
         &  0.5878   & 1/2$^{-}$   &  0.93  & 0 ($s$ wave) &  42.68 &  0.45 & \\
         &  1.0949   & 3/2$^{-}$   &  0.43  & 0 ($s$ wave) &  49.02 &  0.29 & \\  
\hline\hline
\end{tabular}
\end{center}
\end{table*} 

\subsubsection{$^{94}$Mo($\gamma$,n)} 
\indent The highest photon beam energy reached in this experiment at $E_{\gamma}$ = 13.5 MeV will give access to an excitation window in $^{93}$Mo up to $\sim$3.8 MeV from the neutron emission threshold at $S_{n}$ = 9.68 MeV. At photon beam energies $E_{\gamma}$ $<$ $S_{n}$ + 943 keV, the $^{94}$Mo($\gamma$,n)$^{93}$Mo reaction can proceed only to the ground state of $^{93}$Mo. This path would be strongly hampered due to the large angular momentum required for the emitted neutron ($p$ wave) from the compound nucleus $^{94}$Mo$^*$ with $J^{\pi}$ = 1$^{-}$ to the ground state of $^{93}$Mo with $J^{\pi}$ = 5/2$^{+}$. Starting at photon beam energies larger than 10.8 MeV the population of the ground state in $^{93}$Mo proceeds predominantly via emitted neutrons ($s$ wave or $p$ wave) from 22 excited states of $^{93}$Mo. Tables II and III present the energies of the emitted neutrons with their corresponding detection efficiencies and branching ratios as simulated in GEANT4 and calculated in TALYS, respectively. The value of the effective neutron efficiency calculated from Eq. (2) for each of the photon beam energy reached in this experiment is presented in the last columns of the tables.\newline
\setlength{\tabcolsep}{0.4cm}
\begin{table*}[!]
\begin{center}
\caption{\textbf{$^{94}$Mo($\gamma$,n)$^{93}$Mo:} Photon beam energies ($E_{\gamma}$), energy levels in $^{89}$Zr ($E_{i}$), energies of emitted neutrons calculated as $E_{n_{i}} = (\frac{93}{94})(E_{\gamma} - S_{n} -E_{i}$), orbital angular momentum of the emitted neutrons ($l_{i}$), percent neutron detection efficiency at each neutron energy ($\epsilon_{n_{i}}$), neutron branching ratio for each neutron energy ($b_{i}$), and the percent effective neutron detection efficiency, as calculated from Eq. (2), for each photon beam energy ($\epsilon_{n}^{eff}$).}
\begin{tabular}{cccccccc} \hline\hline
$E_{\gamma} $ [MeV] & $E_{i} $ [MeV] &  $J_{i}^{\pi_{i}}$ &  $E_{n_{i}}$ [MeV] &  $l_{i}$ & $\epsilon_{n_{i}}$ [\%] & $b_{i}$ & $\epsilon_{n}^{eff}$ [\%] \\
\hline
 9.7     &  0            & 5/2$^{+}$   &  0.02  & 1 ($p$ wave)  &  53.79 &  1          &  \textbf{53.79}\\
 9.75   &  0            & 5/2$^{+}$   &  0.07  & 1 ($p$ wave)  &  53.29 &  1          &  \textbf{53.29}\\
 9.8     &  0            & 5/2$^{+}$   &  0.12  & 1 ($p$ wave)  &  53.03 &  1          &  \textbf{53.03}\\
 9.85   &  0            & 5/2$^{+}$   &  0.17  & 1 ($p$ wave)  &  52.44 &  1          &  \textbf{52.44}\\ 
 9.95   &  0            & 5/2$^{+}$   &  0.27  & 1 ($p$ wave)  &  51.45 &  1          &  \textbf{51.45}\\
 10      &  0            & 5/2$^{+}$   &  0.32  & 1 ($p$ wave)  &  50.30 &  1          &  \textbf{50.30}\\ 
 10.2   &  0            & 5/2$^{+}$   &  0.52  & 1 ($p$ wave)  &  47.74 &  1          &  \textbf{47.74}\\ 
 10.5   &  0            & 5/2$^{+}$   &  0.81  & 1 ($p$ wave)  &  44.03 &  1          &  \textbf{44.03}\\
           &                &                   &           &                       &            &              & \\
 10.8   &  0            & 5/2$^{+}$   &  1.11  & 1 ($p$ wave)  &  40.51 &  0.59      &  \textbf{45.35}\\ 
           &  0.9433   & 1/2$^{+}$   &  0.18  & 1 ($p$ wave) &  52.32 &  0.41      & \\    
           &                &                   &           &                       &            &               & \\
 11      &  0            & 5/2$^{+}$   &  1.31  & 1 ($p$ wave)  &  38.37 &  0.46      &  \textbf{44.73}\\
           &  0.9433   & 1/2$^{+}$   &  0.37  & 1 ($p$ wave) &  50.15 &  0.54      & \\
           &                &                   &           &                       &            &               & \\             
 11.5   &  0            & 5/2$^{+}$   &  1.80  & 1 ($p$ wave)  &  34.35 &  0.31      &  \textbf{42.83}\\ 
           &  0.9433   & 1/2$^{+}$   &  0.87  & 1 ($p$ wave) &  43.23 &  0.37      & \\
           &  1.4925   & 3/2$^{+}$   &  0.33  & 1 ($p$ wave) &  50.19 &  0.26      & \\
           &  1.6950   & 5/2$^{+}$   &  0.13  & 1 ($p$ wave) &  52.18 &  0.06      & \\
           &                &                   &           &                        &           &               & \\
 11.65 &  0            & 5/2$^{+}$   &  1.95  & 1 ($p$ wave)   &  33.14 &  0.29      &  \textbf{42.46}\\
           &  0.9433   & 1/2$^{+}$   &  1.02  & 1 ($p$ wave)  &  41.68 &  0.33      & \\
           &  1.4925   & 3/2$^{+}$   &  0.47  & 1 ($p$ wave)  &  48.39 &  0.28      & \\   
           &  1.6950   & 5/2$^{+}$   &  0.27  & 1 ($p$ wave)  &  50.43 &  0.10      & \\
           &                &                   &           &                        &            &               & \\
  11.8  &  0            & 5/2$^{+}$   &  2.10  & 1 ($p$ wave)   &  32.25 &  0.29      &  \textbf{40.71}\\
           &  0.9433   & 1/2$^{+}$   &  1.17  & 1 ($p$ wave)  &  39.94 &  0.30      & \\
           &  1.4925   & 3/2$^{+}$   &  0.62  & 1 ($p$ wave)  &  46.42 &  0.28      & \\   
           &  1.6950   & 5/2$^{+}$   &  0.42  & 1 ($p$ wave)  &  49.04 &  0.13      & \\
           &                &                   &           &                        &            &               & \\
11.95  &  0            & 5/2$^{+}$   &  2.25  & 1 ($p$ wave)   &  32.99 &  0.26      &  \textbf{41.57}\\
           &  0.9433   & 1/2$^{+}$   &  1.31  & 1 ($p$ wave)  &  38.36 &  0.25      & \\
           &  1.4925   & 3/2$^{+}$   &  0.77  & 1 ($p$ wave)  &  44.66 &  0.24      & \\   
           &  1.6950   & 5/2$^{+}$   &  0.57  & 1 ($p$ wave)  &  47.25 &  0.11      & \\
           &  2.1420   & 5/2$^{+}$   &  0.129  & 1 ($p$ wave)  &  52.30 & 0.04     & \\
           &  2.1454   & 3/2$^{+}$,  5/2$^{+}$  &  0.125  & 1 ($p$ wave)  &  52.32 & 0.07      & \\   
           &  2.1811   & 3/2$^{+}$   &  0.09    & 1 ($p$ wave)  &  52.83 &  0.03      & \\                  
           &                &                   &             &                        &            &               & \\         
12.25  &  0            & 5/2$^{+}$   &  2.25  & 1 ($p$ wave)   &  29.54 &  0.23      &  \textbf{41.32}\\
           &  0.9433   & 1/2$^{+}$   &  1.61  & 1 ($p$ wave)  &  35.57 &  0.16      & \\
           &  1.4925   & 3/2$^{+}$   &  1.07  & 1 ($p$ wave)  &  40.99 &  0.17      & \\   
           &  1.6950   & 5/2$^{+}$   &  0.87  & 1 ($p$ wave)  &  43.44 &  0.08      & \\
           &  2.1420   & 5/2$^{+}$   &  0.43  & 1 ($p$ wave)  &  48.93 &  0.07      & \\
           &  2.1454   & 3/2$^{+}$,  5/2$^{+}$  &  0.42  & 1 ($p$ wave)  &  48.92 &  0.12      & \\   
           &  2.1811   & 3/2$^{+}$   &  0.39    & 1 ($p$ wave)  &  49.72 &  0.12      & \\      
           &  2.4374   & 1/2$^{+}$   &  0.13  & 1 ($p$ wave)  &  51.98 &  0.04     & \\   
           &  2.5297   & 1/2$^{-}$,  3/2$^{-}$    &  0.04    & 0 ($s$ wave)  &  52.82 &  0.01      & \\                                                
\hline\hline
\end{tabular}
\end{center}
\end{table*} 
\setlength{\tabcolsep}{0.4cm}
\begin{table*}[!]
\begin{center}
\caption{\textbf{$^{94}$Mo($\gamma$,n)$^{93}$Mo:} Photon beam energies ($E_{\gamma}$), energy levels in $^{89}$Zr ($E_{i}$), energies of emitted neutrons calculated as $E_{n_{i}} = (\frac{93}{94})(E_{\gamma} - S_{n} -E_{i}$), orbital angular momentum of the emitted neutrons ($l_{i}$), percent neutron detection efficiency at each neutron energy ($\epsilon_{n_{i}}$), neutron branching ratio for each neutron energy ($b_{i}$), and the percent effective neutron detection efficiency, as calculated from Eq. (2), for each photon beam energy ($\epsilon_{n}^{eff}$).}
\begin{tabular}{cccccccc} \hline\hline
$E_{\gamma} $ [MeV] & $E_{i} $ [MeV] &  $J_{i}^{\pi_{i}}$ &  $E_{n_{i}}$ [MeV] &  $l_{i}$ & $\epsilon_{n_{i}}$ [\%] & $b_{i}$ & $\epsilon_{n}^{eff}$ [\%] \\
\hline  
12.5    &  0            & 5/2$^{+}$   &  2.79  & 1 ($p$ wave)   &  28.78 &  0.23      &  \textbf{39.86}\\
           &  0.9433   & 1/2$^{+}$   &  1.86  & 1 ($p$ wave)  &  33.87 &  0.10      & \\
           &  1.4925   & 3/2$^{+}$   &  1.32  & 1 ($p$ wave)  &  38.37 &  0.19      & \\   
           &  1.6950   & 5/2$^{+}$   &  1.12  & 1 ($p$ wave)  &  40.45 &  0.06      & \\
           &  2.1420   & 5/2$^{+}$   &  0.673  & 1 ($p$ wave)  &  45.74 &  0.05      & \\
           &  2.1454   & 3/2$^{+}$,  5/2$^{+}$  &  0.669  & 1 ($p$ wave)  &  45.60 &  0.10      & \\   
           &  2.1811   & 3/2$^{+}$   &  0.63    & 1 ($p$ wave)  &  46.26 &  0.10      & \\      
           &  2.4374   & 1/2$^{+}$   &  0.38  & 1 ($p$ wave)  &  49.81 &  0.08     & \\   
           &  2.5297   & 1/2$^{-}$,  3/2$^{-}$ &  0.29    & 0 ($s$ wave)  &  50.60 &  0.01      & \\  
           &  2.6190   & 1/2$^{-}$,  3/2$^{-}$ &  0.20    & 0 ($s$ wave)  &  51.55 &  0.01      & \\      
           &  2.6701   & 1/2$^{+}$   &  0.15  & 1 ($p$ wave)  &  52.15 &  0.04     & \\   
           &  2.7046   & 1/2$^{+}$   &  0.12  & 1 ($p$ wave)  &  52.26 &  0.03      & \\   
           &                &                   &           &                        &            &               & \\   
  12.8  &  0            & 5/2$^{+}$   &  3.09  & 1 ($p$ wave)   &  27.66 &  0.28440  &  \textbf{37.16}\\
           &  0.9433   & 1/2$^{+}$   &  2.16  & 1 ($p$ wave)  &  31.84 &   0.07408  & \\
           &  1.4925   & 3/2$^{+}$   &  1.61  & 1 ($p$ wave)  &  35.74 &   0.14420  & \\   
           &  1.6950   & 5/2$^{+}$   &  1.41  & 1 ($p$ wave)  &  37.65 &   0.07771  & \\
           &  2.1420   & 5/2$^{+}$   &  0.970  & 1 ($p$ wave)  &  42.27 & 0.03704  & \\
           &  2.1454   & 3/2$^{+}$,  5/2$^{+}$  &  0.966  & 1 ($p$ wave)  &  42.25 & 0.07167 & \\   
           &  2.1811   & 3/2$^{+}$   &  0.93    & 1 ($p$ wave)  &  46.46 &  0.06981 & \\      
           &  2.4374   & 1/2$^{+}$   &  0.68  & 1 ($p$ wave)  &  45.92 &    0.06373 & \\   
           &  2.5297   & 1/2$^{-}$,  3/2$^{-}$ &  0.59    & 0 ($s$ wave)  &  46.71 &  0.01074  & \\  
           &  2.6190   & 1/2$^{-}$,  3/2$^{-}$ &  0.50    & 0 ($s$ wave)  &  48.01 &  0.00883   & \\      
           &  2.6701   & 1/2$^{+}$   &  0.45  & 1 ($p$ wave)  &  48.90 &  0.05549 & \\   
           &  2.7046   & 1/2$^{+}$   &  0.41  & 1 ($p$ wave)  &  49.35 &  0.05335 & \\  
           &  2.8421   & 1/2$^{+}$ &  0.28    & 1 ($p$ wave)  &  50.66 &  0.04113 & \\      
           &  2.9552   & 1/2$^{-}$,  3/2$^{-}$  &  0.17  & 0 ($s$ wave)  &  51.80 & 0.00499 & \\   
           &  3.0640   & 1/2$^{-}$,  3/2$^{-}$  &  0.06  & 0 ($s$ wave)  &  52.66 & 0.00283 & \\                    
           &                &                   &           &                        &            &               & \\              
13.5    &  0            & 5/2$^{+}$   &  3.78  & 1 ($p$ wave)   &  25.35 & 0.32964 &  \textbf{32.93}\\
           &  0.9433   & 1/2$^{+}$   &  2.85  & 1 ($p$ wave)  &  28.48 &  0.07525 & \\
           &  1.4925   & 3/2$^{+}$   &  2.30  & 1 ($p$ wave)  &  30.79 &  0.10762 & \\   
           &  1.6950   & 5/2$^{+}$   &  2.10  & 1 ($p$ wave)  &  32.17 &  0.07133 & \\
           &  2.1420   & 5/2$^{+}$   &  1.660  & 1 ($p$ wave)  &  35.24 & 0.03091 & \\
           &  2.1454   & 3/2$^{+}$,  5/2$^{+}$  &  1.659  & 1 ($p$ wave)  &  35.27 & 0.05072 & \\   
           &  2.1811   & 3/2$^{+}$   &  1.62    & 1 ($p$ wave)  &  35.53 &  0.04801 & \\      
           &  2.4374   & 1/2$^{+}$   &  1.37  & 1 ($p$ wave)  &  38.07 &  0.04383 & \\   
           &  2.5297   & 1/2$^{-}$,  3/2$^{-}$ &  1.28    & 0 ($s$ wave)  &  38.71 &  0.01244 & \\  
           &  2.6190   & 1/2$^{-}$,  3/2$^{-}$ &  1.19    & 0 ($s$ wave)  &  39.69 &  0.00935 & \\      
           &  2.6701   & 1/2$^{+}$   &  1.14  & 1 ($p$ wave)  &  40.15 &  0.04146 & \\   
           &  2.7046   & 1/2$^{+}$   &  1.11  & 1 ($p$ wave)  &  40.43 &  0.04142 & \\  
           &  2.8421   & 1/2$^{+}$ &  0.97    & 1 ($p$ wave)  &  42.03 &  0.04114 & \\      
           &  2.9552   & 1/2$^{-}$,  3/2$^{-}$  &  0.86  & 0 ($s$ wave)  &  43.52 &  0.00869 & \\   
           &  3.0640   & 1/2$^{-}$,  3/2$^{-}$  &  0.75  & 0 ($s$ wave)  &  44.92 &  0.00680 & \\        
           &  3.1592   & 3/2$^{+}$,  5/2$^{+}$  &  0.66  & 1 ($p$ wave)  &  45.77 & 0.02017 & \\              
           &  3.3876   & 3/2$^{+}$,  5/2$^{+}$  &  0.43  & 1 ($p$ wave)  &  48.92 &  0.01743 & \\              
           &  3.4503   & 3/2$^{+}$,  5/2$^{+}$  &  0.37  & 1 ($p$ wave)  &  49.78 &  0.03078 & \\               
           &  3.5900   & 1/2$^{-}$,  3/2$^{-}$  &  0.23  & 0 ($s$ wave)  &  50.66 &  0.00354 & \\  
           &  3.5963   & 3/2$^{+}$,  5/2$^{+}$  & 0.22  & 1 ($p$ wave)  & 51.08 &  0.00348 & \\   
           &  3.7089   & 3/2$^{+}$,  5/2$^{+}$  & 0.11  & 1 ($p$ wave)  & 52.06 &  0.00241  & \\                          
           &  3.7200   & 1/2$^{-}$,  3/2$^{-}$  &  0.10  & 0 ($s$ wave)  &  52.33 &  0.00229  & \\  
           &  3.7900   & 1/2$^{-}$,  3/2$^{-}$  &  0.03  & 0 ($s$ wave)  &  52.57 &  0.00129  & \\                                                        
\hline\hline
\end{tabular}
\end{center}
\end{table*} 
To assess the uncertainty in the calculation of the neutron branching ratios in TALYS we considered another four sets of nuclear inputs such as:

\paragraph{INPUT-1} {\textbf{Level density:} Constant-temperature (CT) plus Fermi gas model {\cite{Koni08}; \textbf{Optical potential:} Koning and Delaroche \cite{Koni03}; \textbf{$\gamma$ strength:} axially-symmetric-deformed Hartree-Fock-Bogoliubov (HFB) plus QRPA model based on the D1M Gogny interaction \cite{Mart16,Gori16,Peru08,Gori18}
\paragraph{INPUT-2} {\textbf{Level density:} HFB plus combinatorial nuclear level model \cite{Gori08}; \textbf{Optical potential:} Koning and Delaroche \cite{Koni03}; \textbf{$\gamma$ strength:} Generalized Lorentzian (GLO) model \cite{Kope90,Capo09}
\paragraph{INPUT-3} {\textbf{Level density:} HFB plus combinatorial nuclear level model \cite{Gori08}; \textbf{Optical potential:} semi-microscopic neutron-nucleus spherical optical model potential from the nuclear matter approach of Jeukenne, Lejeune, and Mahaux (JLM) \cite{Baug98,Baug01}; \textbf{$\gamma$ strength:} axially-symmetric-deformed Hartree-Fock-Bogoliubov (HFB) plus QRPA model based on the D1M Gogny interaction \cite{Mart16,Gori16,Peru08,Gori18}
\paragraph{INPUT-4} {\textbf{Level density:} HFB plus combinatorial nuclear level model \cite{Gori08}; \textbf{Optical potential:} semi-microscopic neutron-nucleus spherical optical model potential from the nuclear matter approach of Jeukenne, Lejeune, and Mahaux (JLM) \cite{Baug98,Baug01}; \textbf{$\gamma$ strength:} Generalized Lorentzian (GLO) model \cite{Kope90,Capo09}.\newline
We obtained very similar values for the neutron branching ratios with differences varying between less than 1$\%$ and $\sim$3$\%$.

\subsection{Beam-induced background measurements}

\indent Accurate neutron counting in the $^{3}$He counters requires the ability to distinguish ($\gamma$,n) events originating in the target from beam-induced background events. That is particularly important for the ($\gamma$,n) reaction cross sections measured at the neutron emission threshold where the photoneutron cross sections are very small but astrophysically relevant.\newline
\indent For the $^{94}$Mo($\gamma$,n) reaction cross section measurements, the $^{90}$Zr target with an atomic mass number $Z$ = 40, close to the atomic mass number of $Z$ = 42 for Mo but with a higher $S_{n}$ of 11.97 MeV, was used to mimic the $\gamma$-ray induced background in the $^{3}$He counters caused by Compton scattering and pair production from the $^{94}$Mo target. Because $^{nat}$H has a deuterium ($S_{n}$ = 2.225 MeV) abundance of 0.016\% and the d($\gamma$,n) reaction cross section peaks at 2.5 mb, a significant amount of H needs to be in the path of the $\gamma$-ray beam for this background to be measurable. The polyethylene moderator of the INVS, however, cannot be removed as it is an integral part of the detector. Thus, $\gamma$-ray beam induced-neutron background measurements with the $^{90}$Zr target were carried out at photon beam energies corresponding to the cross section measurements of the $^{94}$Mo($\gamma$,n) reaction. At the neutron emission threshold ($S_{n}$ = 9.68 MeV), the background was about 25\% of the total counting rate of the INVS detector. Once the rates of the beam-induced background were measured on the $^{90}$Zr target, the corresponding background rates for the $^{94}$Mo($\gamma$,n) reaction cross section measurements were determined by scaling the rates with a factor of $\sim$0.6 which comes from target thickness normalization.\newline
\indent For the $^{90}$Zr($\gamma$,n) reaction cross section measurements, the beam-induced background rates registered in the INVS detector at $\gamma$-ray beam energies below the $^{90}$Zr neutron emission threshold were extrapolated by a linear fit to the energies above the threshold at which the $^{90}$Zr($\gamma$,n) reaction cross section measurements were performed.\newline
\indent In addition to the beam-induced background runs, empty target runs were also carried out to account for possible neutron-induced background by the INVS detector itself. At $\gamma$-ray beam energies close to the neutron threshold, the empty target rates, of about 3\% of the total INVS rate, were insignificant compared to the rates recorded for the $\gamma$-ray-beam-induced background on the $^{90}$Zr target.

\section{Data analysis and results}

\indent Under the assumption of a monoenergetic $\gamma$-ray beam, the photoneutron reaction cross section as a function of beam energy may be written as:

\begin{equation}
\sigma_{(\gamma,n)}(E_{\gamma})=(R_{n}-R_{bkgd})/(R_{\gamma} \cdot N_{t} \cdot f \cdot \epsilon^{eff}_{n}(E_{\gamma})),
\end{equation}
where $R_{n}$ is the rate of total number of detected neutrons, $R_{bkgd}$ is the rate of background events, $R_{\gamma}$ is the rate of the incident $\gamma$-ray beam, $N_{t}$ is the number of atoms in the target per unit area, $f$ is the thick-target correction factor calculated as $f = (1-e^{-\mu d})/(\mu d)$ with the linear attenuation coefficient of photons ($\mu$) and the target thickness ($d$), and $\epsilon^{eff}_{n}$ is the effective neutron detection efficiency as calculated from Eq. (2).  $R_{\gamma}$ was determined as the ratio between the rate measured in the paddle detector and the detection efficiency of the paddle detector.\newline
\indent Contributions to the systematic experimental uncertainties of the cross section measurements come mainly from target thickness (2\%), the error in the simulation of the $^{3}$He counter array's neutron detection efficiency (3\%), and the error in the simulation of the efficiency of the paddle detector (4\%).\newline
\indent In Tables IV and V the experimental values are presented for the two ($\gamma$,n) reaction cross sections obtained using Eq. (3). \newline
\setlength{\tabcolsep}{0.2cm}
  \begin{table}[htbp]
     \centering
     \caption{Experimental cross sections for the $^{90}$Zr($\gamma$,n) reaction, as determined from Eq. (3), along with their uncertainties. Measurements were performed for a Gaussian full-energy peak of the $\gamma$-ray beams with mean $E_{\gamma}$ and spread $\sigma_{E_{\gamma}}$. The $\eta$ ratio in the last column is the ratio between the neutron detection efficiency of the ($\gamma,n_{0}$) channel and the effective neutron efficiency, which is also the ratio between the measured $\sigma_{(\gamma,n)}$ cross section and the $\sigma_{(\gamma,n_{0})}$ cross section that would correspond to the detection of neutrons emitted only when the ground state is populated directly. See TABLE I for the corresponding values of $\epsilon_{n_{0}}$ and $\epsilon_{n}^{eff}$.} 
     \begin{tabular} {@{} cccc @{}} 
        \hline\hline
           \\    
             $E_{\gamma} [MeV]$ & $\sigma_{E_{\gamma}} [MeV]$ &$\sigma_{(\gamma,n)} [mb]$ & $\eta$ = $\frac{\epsilon_{n_{0}}}{\epsilon_{n}^{eff}}$ = $\frac{\sigma_{(\gamma,n)}}{\sigma_{(\gamma,n_{0})}}$ \\        
        \\
        \hline
        11.75  &  0.21  & 0.01 $\pm$ 0.01  & 1\\
        12       &  0.23  & 0.11  $\pm$ 0.01  & 1\\
        12.1    &  0.21  & 0.14  $\pm$ 0.02  & 1\\
        12.2    &  0.22  & 0.50  $\pm$ 0.03  & 1\\
        12.4    &  0.22  & 2.28  $\pm$ 0.12  & 1\\
        12.5    &  0.23  & 4.42  $\pm$ 0.24  & 1\\
        12.8    &  0.23  & 9.67  $\pm$ 0.52  & 0.88\\
        13       &  0.22  & 12.66  $\pm$ 0.68  & 0.88\\
        13.5    &  0.24  & 20.94  $\pm$ 1.13  & 0.85\\
        \hline\hline
     \end{tabular}
     \label{tab:booktabs}
  \end{table}
\setlength{\tabcolsep}{0.2cm}
  \begin{table}[htbp]
     \centering
     \caption{Experimental cross sections for the $^{94}$Mo($\gamma$,n) reaction, determined from Eq. (3), along with their uncertainties (same as TABLE IV). See TABLES II and III for the corresponding values of $\epsilon_{n_{0}}$ and $\epsilon_{n}^{eff}$.} 
     \begin{tabular} {@{} cccc @{}} 
        \hline\hline
           \\    
             $E_{\gamma} [MeV]$ & $\sigma_{E_{\gamma}} [MeV]$ & $\sigma_{(\gamma,n)} [mb]$ & $\eta$ = $\frac{\epsilon_{n_{0}}}{\epsilon_{n}^{eff}}$ = $\frac{\sigma_{(\gamma,n)}}{\sigma_{(\gamma,n_{0})}}$\\        
        \\
        \hline
        9.5     &  0.18    & 0.28  $\pm$ 0.02   &1\\
        9.6     &  0.17    & 1.21  $\pm$ 0.07   &1\\
        9.65   &  0.17    & 2.51  $\pm$ 0.14   &1\\
        9.7     &  0.17    & 2.97  $\pm$ 0.16   &1\\
        9.75   &  0.17    & 4.50  $\pm$ 0.24   &1\\
        9.8     &  0.17    & 4.93  $\pm$ 0.27   &1\\
        9.85   &  0.17    & 6.28  $\pm$ 0.34   &1\\
        9.95   &  0.16    & 7.83  $\pm$ 0.42    &1\\
        10      &  0.19    & 8.44  $\pm$ 0.46    &1\\
        10.2   &  0.17    & 10.11 $\pm$ 0.55   &1\\
        10.5   &  0.17    & 11.77  $\pm$ 0.63   &1\\
        10.8   &  0.17    & 13.06  $\pm$ 0.70   &0.89\\
        11      &  0.17    & 14.53  $\pm$ 0.78   &0.86\\
        11.5   &  0.24    & 17.47  $\pm$ 0.94   &0.80\\
        11.65 &  0.25    & 18.73 $\pm$ 1.01   &0.78\\
        11.8   &  0.22    &  20.63  $\pm$ 1.11   &0.79\\
        11.95 &  0.23    &  22.61  $\pm$ 1.22   &0.79\\
        12.25 &  0.22    &  24.20  $\pm$ 1.30  &0.71\\
        12.5   &  0.23     &  27.86  $\pm$ 1.50  &0.72\\
        12.8   &  0.23     &  32.39  $\pm$ 1.74  &0.74\\
        13.5   &  0.24     & 48.64  $\pm$ 2.62  &0.77\\  
        \hline\hline
     \end{tabular}
     \label{tab:booktabs}
  \end{table}
\indent As mentioned previously, the HI$\gamma$S $\gamma$-ray beam had an energy width in the range of 4\% - 5\% (FWHM). Hence, the experimental cross sections determined from Eq. (2) do not represent cross section values at single energies, but rather cross section integrated over an energy range defined by the width of the $\gamma$-ray beam energy profile. The effects of this convolution are particularly significant near the neutron emission threshold, where the cross section changes rapidly and the beam energy distribution extends both above and below the reaction threshold. Since this is the energy region most relevant in astrophysics, the energy window of effective stellar burning for photoneutron reactions, it is important to deconvolve the effects of the finite $\gamma$-ray beam energy distribution to recover the photoneutron cross section. \newline
 \indent An iterative fitting procedure, dubbed ICARUS ($\textbf{I}$terative $\textbf{C}$ode for $\textbf{A}$utomatically $\textbf{R}$esolving and $\textbf{U}$nfolding $\textbf{S}$pectral effects) has been developed for finding an analytical excitation function that when convolved with the $\gamma$-ray beam energy profile will reproduce best the experimental cross sections determined from Eq. (3). ICARUS takes an arbitrary, user defined function with an arbitrary number of fit parameters to represent the photoneutron reaction cross section. ICARUS then convolves that analytical function with the gaussian energy spectrum of the HI$\gamma$S photon beam to produce the effective cross section, or yield, that is measured experimentally. The fit parameters are then varied to minimize the $\chi^2$-value of the ICARUS calculated yields, compared with the experimentally measured yields.\newline
 \indent In the case of the $^{94}$Mo($\gamma$,n) reaction, a Gaussian $\gamma$-ray beam energy profile with experimentally measured mean and width values was convolved with an ICARUS excitation function that was a 8-parameter function described as a product between the threshold behavior of the ($\gamma$,n) reaction cross section from Ref. \cite{Vogh02} and a fifth-degree polynomial function as follows: 
 \begin{equation}
 \begin{split}
 \sigma_{(\gamma,n)}^{ICARUS}(E_{\gamma})=\sigma_{0}[(E_{\gamma}-S_{n})/S_{n}]^{p_{1}}\cdot[p_{2}   \\
 + p_{3}\cdot(E_{\gamma}-S_{n}) \\
 + p_{4}\cdot(E_{\gamma}-S_{n})^2+p_{5}\cdot(E_{\gamma}-S_{n})^3 \\
 + p_{6}\cdot(E_{\gamma}-S_{n})^4+p_{7}\cdot(E_{\gamma}-S_{n})^5].
 \end{split}
 \end{equation}
 The best fit values of the $\sigma_{0}$ and $p_{i}$ (i = 1,7) parameters of the analytical cross section function from Eq. (4) are $-$ 30.59 mb, 0.585, 2.961, -3.215, 2.693, -1.098, 0.226, -0.017, respectively. \newline
 \indent Figure 4 shows the ICARUS fitting results for the experimental excitation function of the $^{94}$Mo($\gamma$,n) reaction. The ICARUS fitting procedure was not applicable in case of the $^{90}$Zr($\gamma$,n) reaction due to the scarcity of the experimental data points. 
\begin{figure}[h]
\includegraphics[scale=0.3,keepaspectratio=true]{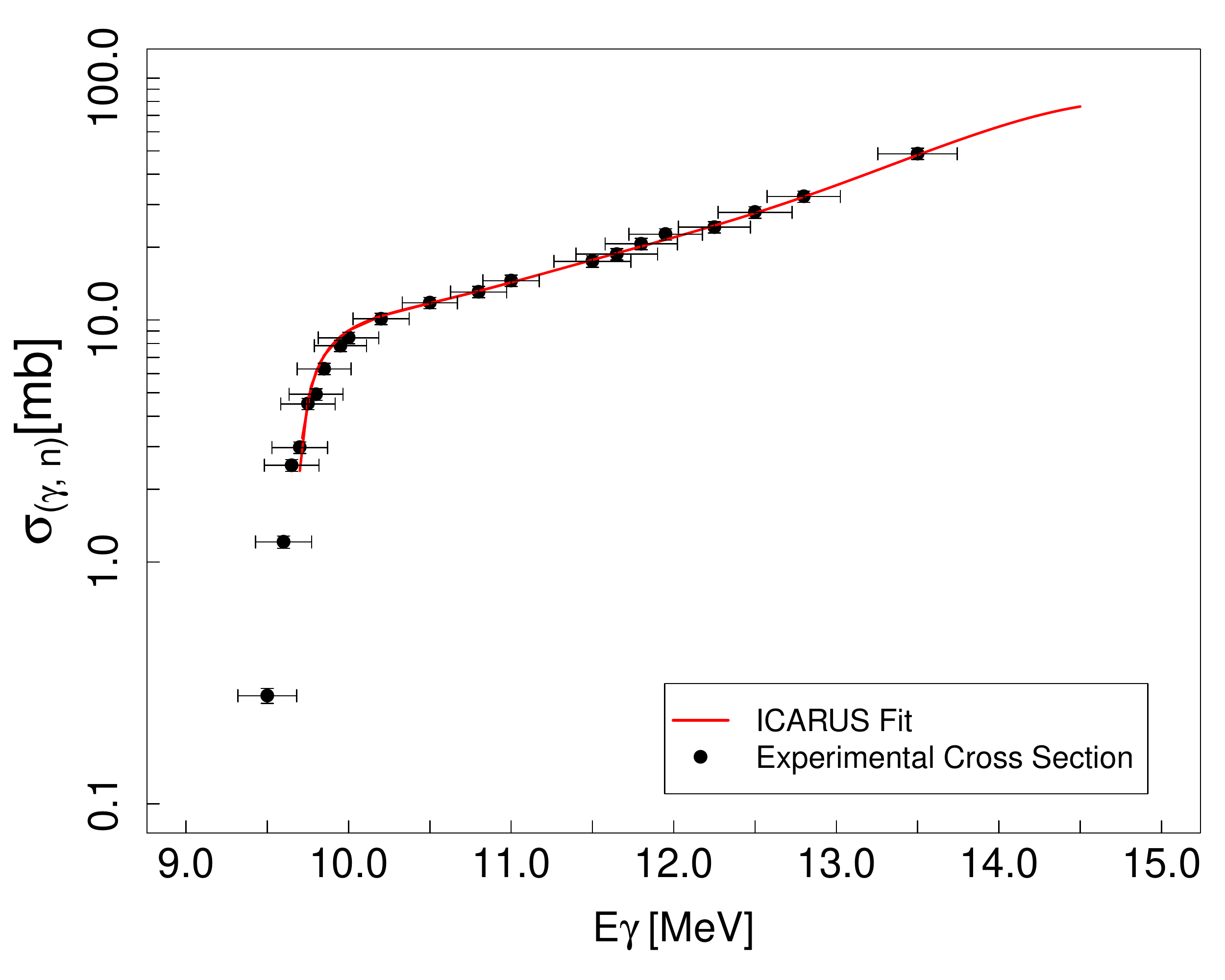}
\caption[Fig. 4]{(Color online) ICARUS fitting plot for the excitation function of the $^{94}$Mo($\gamma$,n) reaction. The horizontal error bars represent the measured $\gamma$-ray beam energy widths.}
\end{figure}    

\indent The \textit{ICARUS excitation function for the $^{94}$Mo($\gamma$,n) reaction} corresponding to the experimental $\gamma$-ray energies and \textit{the experimental excitation function for $^{90}$Zr($\gamma$,n) reaction determined from Eq. (3)} are shown in Figure 5 and 6, respectively, in comparison with the previous measurements carried out with quasi-monochromatic laser-Compton scattering photons \cite{Utsu13} and with quasi-monochromatic annihilation photons \cite{Beil74,Berm67,Lepr71}. 
\begin{figure}[h]
\includegraphics[scale=0.3,keepaspectratio=true]{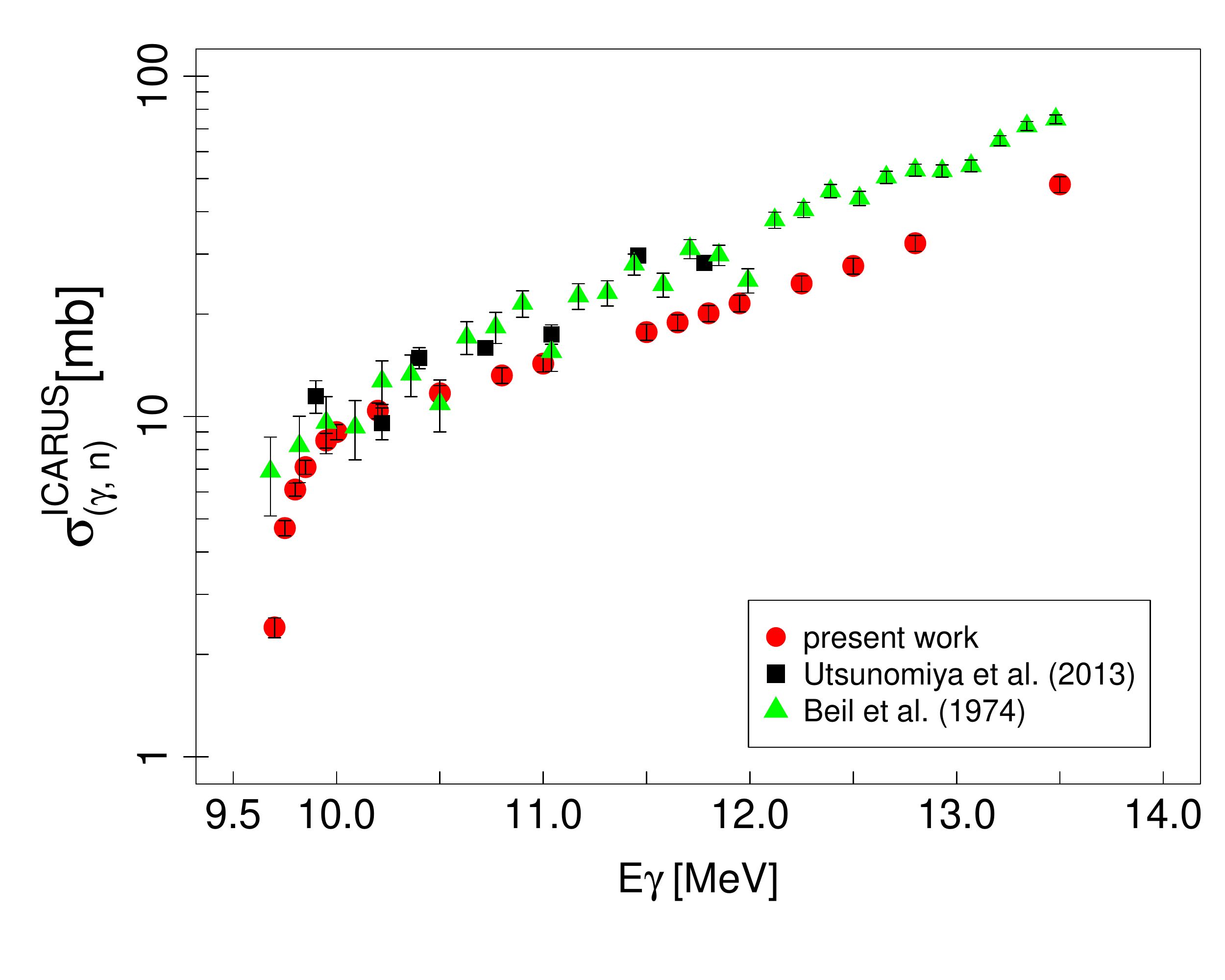}
\caption[Fig. 5]{(Color online) ICARUS excitation function for $^{94}$Mo($\gamma$,n) of this work compared with the previous measurements \cite{Utsu13,Beil74}.}
\end{figure}  

\begin{figure}[h]
\includegraphics[scale=0.3,keepaspectratio=true]{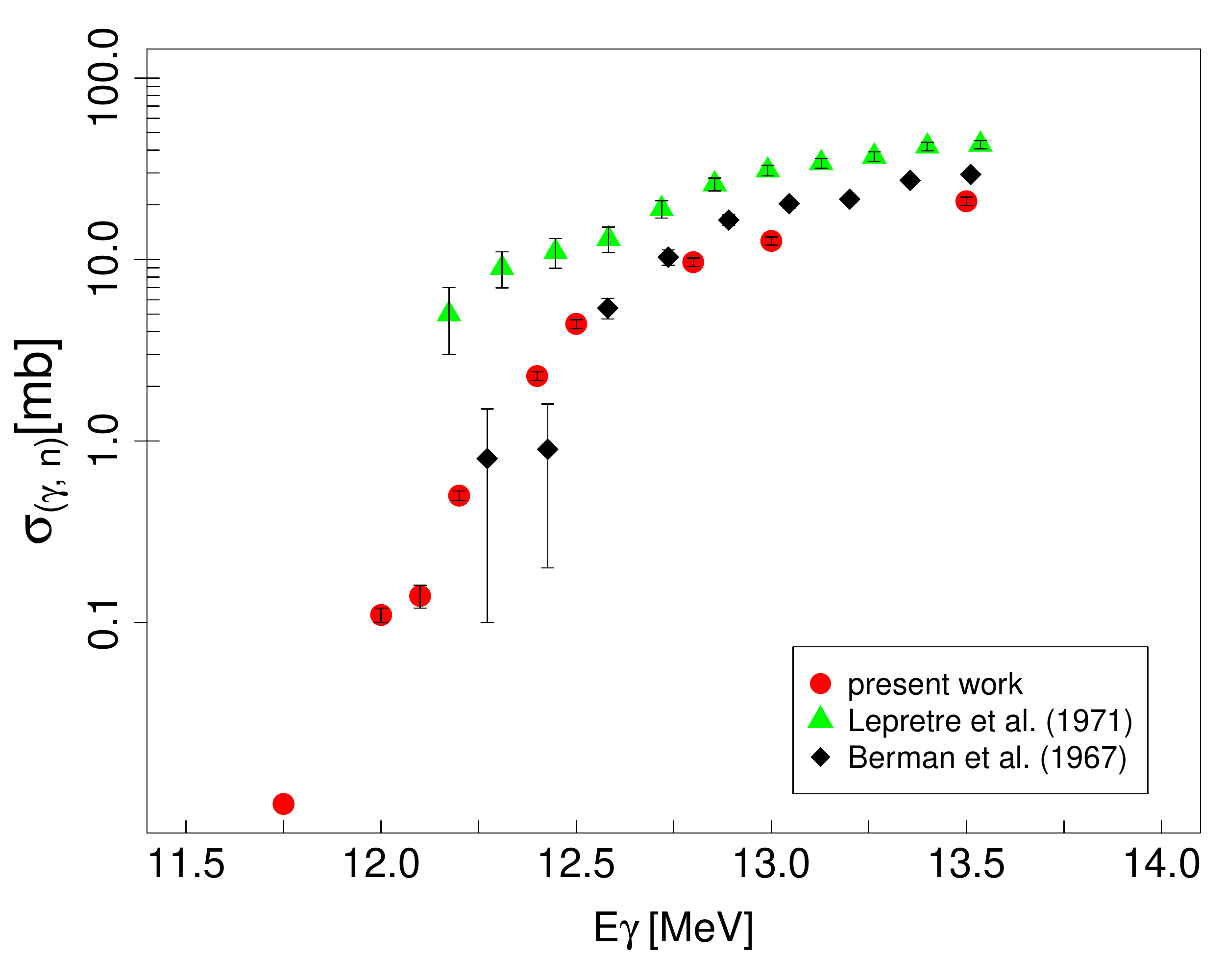}
\caption[Fig. 6]{(Color online) Excitation function for $^{90}$Zr($\gamma$,n) of this work compared with the previous measurements \cite{Berm67,Lepr71}.}
\end{figure}  

\indent In the case of the $^{94}$Mo($\gamma$,n) reaction, Fig. 5 shows that there is a good agreement between the present results and the previous results by Ustunomiya $et $ $al.$ \cite{Utsu13} and by Beil $et$ $al.$ \cite{Beil74} for photon beam energies below 10.8 MeV. However above that energy, when neutrons emitted from excited states in $^{93}$Mo that $\gamma$ decay to the ground state contribute to the measured cross sections, present results start to deviate from previous work. Similarly, in the case of the $^{90}$Zr($\gamma$,n) reaction, Fig. 6 shows that the present results are in good agreement with the results by Berman $et$ $al.$ \cite{Berm67} for photon beam energies below 12.8 MeV, whereas above 12.8 MeV when neutrons emitted from excited states in $^{89}$Zr that $\gamma$ decay to the ground state contribute to the measured cross sections, the agreement worsens. However, the results of Lepr$\hat{e}$tre $et$ $al.$ \cite{Lepr71} show outstanding discrepancies both with our results and with the results of Ref. \cite{Berm67} for all photon beam energies. Berman $et$ $al.$ \cite{Berm87} reviewed the inconsistencies between the data of Refs. \cite{Berm67} and \cite{Lepr71}, notable not only for $^{90}$Zr but also for a few other cases where results disagree in the GDR peak height by 15\% or more. Particularly for $^{90}$Zr, the previous results of Ref. \cite{Berm67} were confirmed, which led to the conclusion in Ref. \cite{Berm87} that there was an error for the dataset of Ref. \cite{Lepr71} either in the photon flux determination or in the neutron detection efficiency or in both. \newline
\indent In a more recent data evaluation, Varlamov $et$ $al.$ \cite{Varl09} stated that the incorrectness of the special procedure used in the experiments of Ref. \cite{Lepr71} to sort photoneutrons in multiplicity is the reason behind the observed discrepancies. This reasoning is used as well in the current EXFOR database \cite{Exfo}. \newline
\indent Note that the discrepancies between the data of Refs. \cite{Berm67} and \cite{Lepr71} are currently under review within the framework of the IAEA Coordinated Research Project entitled ``Updating the Photoneutron Data Library and Generating a Reference Database for Photon Strength Functions".

\section{Statistical model calculations}

\indent The photoneutron reaction cross sections of the present work are now compared with theoretical calculations obtained with the TALYS nuclear reaction code \cite{Koni12} and two different models of the $\gamma$SF, namely the Generalized Lorentzian (GLO) model \cite{Kope90,Capo09} and the axially-symmetric-deformed Hartree-Fock-Bogoliubov (HFB) plus QRPA model based on the D1M Gogny interaction \cite{Mart16,Gori16,Peru08,Gori18}. The D1M$+$QRPA model includes phenomenologically the impact of multi-particle multi-hole configurations as well as phonon coupling and has proven its capacity to reproduce experimental data relatively well \cite{Mart16,Gori16,Gori18}. Both the GLO and D1M+QRPA models are standard inputs in TALYS reaction code and are classically used for practical applications. Since they are based on fundamentally different physics, they can reflect the existing uncertainties affecting the $\gamma$SF, but also the impact of such uncertainties on reaction cross sections and astrophysical rates. The HFB plus combinatorial nuclear level density model \cite{Gori08} is used for the present photoneutron reaction cross section calculations.

\subsection{Cross section calculations and comparison with experimental results}

\indent The D1M$+$QRPA calculation has been renormalized, as detailed in Refs. \cite{Mart16,Gori16,Gori18}, in order to reproduce the present data. As seen in Fig. 7, this leads, however, to some overestimate of the data in the vicinity of 13 MeV for the $^{90}$Zr($\gamma$,n) reaction, while the traditional GLO model, adjusted to the former data \cite{Berm67,Lepr71}, strongly overpredicts the present data in the vicinity of 13 MeV and above. The large GDR width adopted in the GLO model to reproduce the data of Ref. \cite{Lepr71} in the 13-15 MeV region may be questionable, especially in view of our new low-energy measurements below 14 MeV.
\begin{figure}[h]
\includegraphics[scale=0.35,keepaspectratio=true]{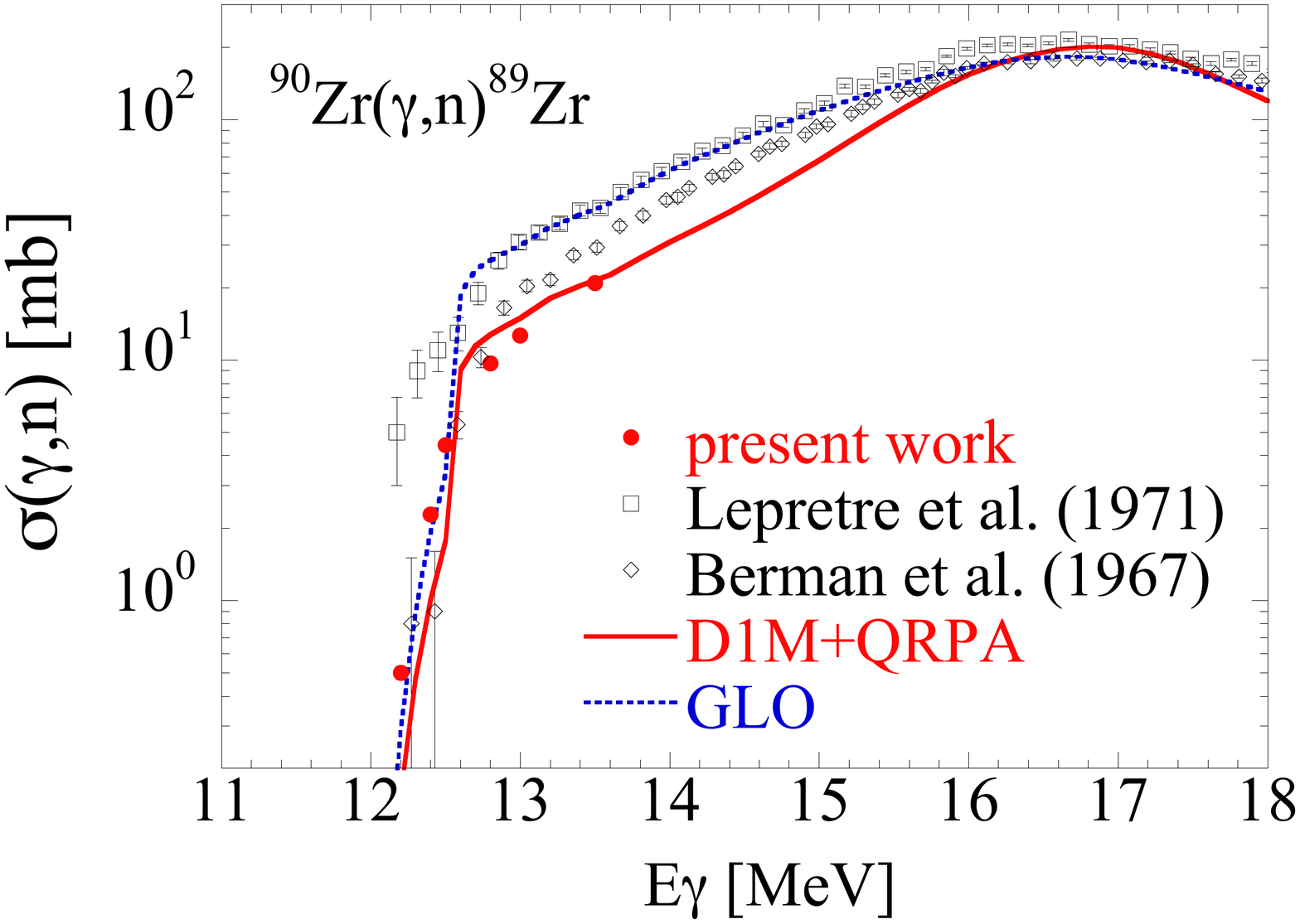}
\caption[Fig. 7]{(Color online) Comparison between the present ($\gamma$,n) photoneutron reaction cross sections as a function of the $\gamma$-ray beam energy and the previous data \cite{Berm67,Lepr71} for $^{90}$Zr. Also included are the predictions obtained with the D1M$+$QRPA $E1$ and $M1$ strengths (solid line) and with the GLO model (dotted line).}
\end{figure}  

\indent Good agreement between experimental and theoretical photoneutron reaction cross sections is obtained for the $^{94}$Mo($\gamma$,n) reaction. In particular, the present data agree fairly well with previous measurements \cite{Utsu13,Beil74} in the 10-11 MeV range. Fig. 8 shows both the GLO and D1M$+$QRPA models adjusted to reproduce experimental photoneutron cross sections.
\begin{figure}[h]
\includegraphics[scale=0.35,keepaspectratio=true]{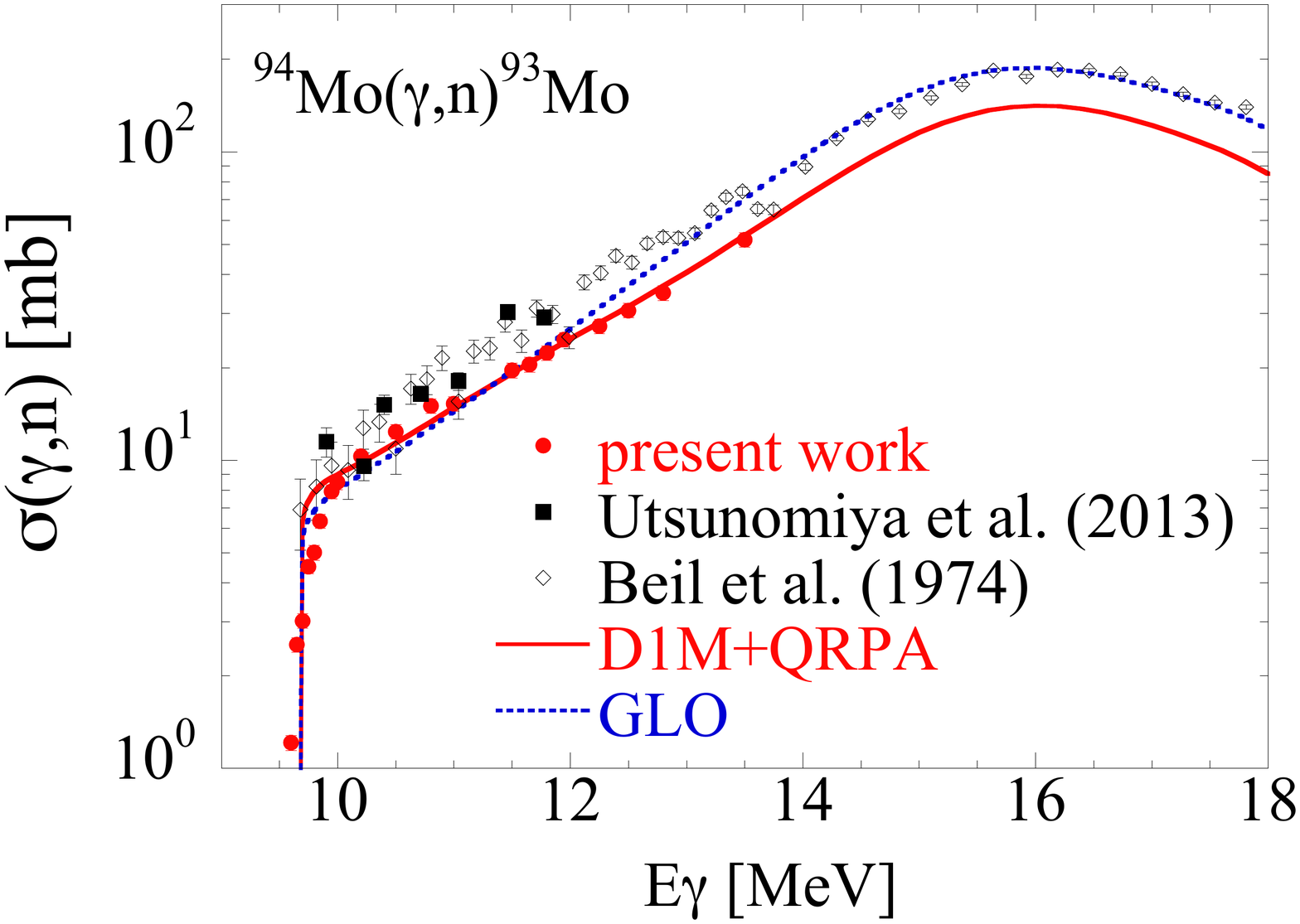}
\caption[Fig. 8]{(Color online) Comparison between the present ($\gamma$,n) photoneutron reaction cross sections as a function of the $\gamma$-ray beam energy and the previous data \cite{Utsu13,Beil74} for $^{94}$Mo. Also included are the predictions obtained with the D1M$+$QRPA $E1$ and $M1$ strengths (solid line) and with the GLO model (dotted line).}
\end{figure}  

In Fig.~\ref{Fig9}, we combine experimental information on the $\gamma$SF below and above the neutron separation energy and compare them with the D1M+QRPA and GLO $\gamma$SF. The $\gamma$SF is connected to the photoneutron cross section through
\begin{equation}
f(E_\gamma)=\frac{1}{3\pi^2\hbar^2c^2}\frac{\sigma_\gamma(E_\gamma)}{E_\gamma},
\label{eq_gsf}
\end{equation}
where the constant $1/3\pi^2\hbar^2c^2=8.67\times10^{-8} {\rm mb^{-1}MeV^{-2}}$. Note that Eq.~\ref{eq_gsf} holds only when the neutron emission channel dominates over the electromagnetic de-excitation, {\it i.e.} a few hundred keV above the neutron threshold. For this reason, photodata in this range have not been included in Fig.~\ref{Fig9}.

Both the $E1$ and $M1$ theoretical components are shown separately, as well as the total dipole $E1+M1$ strength. Although, the spin-flip $M1$ strengths are seen to be quite different, the total $\gamma$SF do not differ significantly, {\it i.e.} typically within less than a factor of 2, even below the neutron separation energy. None of the models are able to reproduce the strong $^{90}$Zr and $^{94}$Mo low-lying strengths obtained by ($\gamma$,$\gamma^\prime$) bremsstrahlung experiments \cite{Schw08,Ruse09}. In contrast, the $^{94}$Mo dipole strength below 8~MeV extracted with the Oslo method from neutron pickup ($^3$He,$\alpha \gamma$) and inelastic scattering ($^3$He,$^3$He$^\prime$) reactions \cite{Utsu13,Gutt05} agrees relatively well with theory, in particular with D1M+QRPA. Note that the experimental data below 3~MeV is associated with the $M1$ de-excitation strength that is not included here, neither in the GLO nor in the D1M+QRPA photoabsorption description \cite{Gori18}. Discrepancies between the bremsstrahlung and Oslo data are still being investigated, in particular within the above-mentioned IAEA Coordinated Research Project.

\begin{figure}[h]
\includegraphics[scale=0.4,keepaspectratio=true]{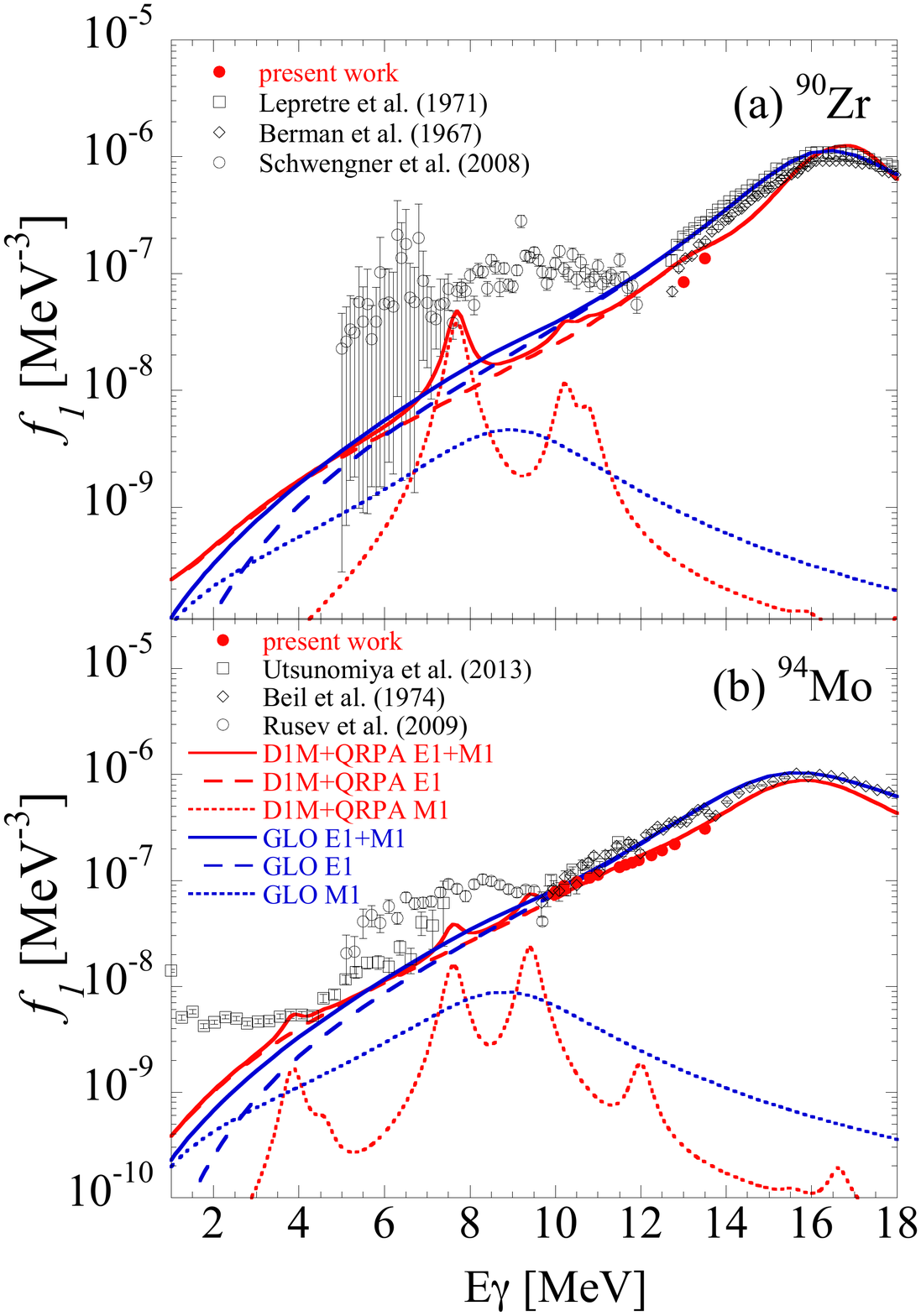}
\caption[Fig. 9]{
(Color online) Comparison between experimental \cite{Berm67,Lepr71,Utsu13,Beil74,Schw08,Ruse09,Gutt05} and theoretical $\gamma$-ray strength function as a function of the $\gamma$-ray energy for (a) $^{90}$Zr and (b) $^{94}$Mo. The predictions correspond to the D1M$+$QRPA $E1$ (red dashed line), $M1$ (red dotted line) and $E1+M1$ (red solid line) strengths and the GLO $E1$ (blue dashed line), $M1$ (blue dotted line) and $E1+M1$ (blue solid line) strengths.}
\label{Fig9}
\end{figure}  

\subsection{Stellar reaction rate calculations}

\indent Nucleosynthesis investigations require the use of stellar rates for thermal population of excited states in the target. Stellar photoneutron reaction rates are calculated in the TALYS code from the expression

 \begin{equation}
 \lambda_{(\gamma,n)}^{\ast}(T)=\frac{\displaystyle \sum_{\mu}(2J^{\mu}+1)\lambda_{(\gamma,n)}^{\mu}(T)exp(-E^{\mu}/kT)}{\displaystyle \sum_{\mu}(2J^{\mu}+1)exp(-E^{\mu}/kT)},
 \end{equation}
where $J_{\mu}$ represents the levels of the target nucleus, $\mu$ labels the thermally populated state, and $E^{\mu}$ stands for the excitation energy of that state. \newline

Photoneutron rates $\lambda_{(\gamma,n)}^{\mu}$(T)  for individual states are found from the integral of a Planck black-body spectrum $n(E_{\gamma},T)$, which describes the energy distribution of the stellar photons, and the associated photoneutron emission cross section

 \begin{equation}
 \lambda_{(\gamma,n)}^{\mu}(T)=\int_{0}^{\infty}cn_{\gamma}(E,T)\sigma_{(\gamma,n)}^{\mu}(E)dE,
 \end{equation}
where $c$ is the speed of light.\newline

\indent In Fig.~\ref{Fig10} are shown the resulting $^{90}$Zr($\gamma$,n)$^{89}$Zr and $^{94}$Mo($\gamma$,n)$^{93}$Mo stellar photoneutron rates as a function of the temperature in a typical range of interest for the $p$-process nucleosynthesis \cite{Arno03}. 

Also shown in  Fig.~\ref{Fig10}, are the competing $^{90}$Zr($\gamma$,p)$^{89}$Y and $^{94}$Mo($\gamma$,$\alpha$)$^{90}$Zr stellar photoreaction rates which dominate the photoneutron channel at temperatures below $T\simeq 4 \times 10^9$K and $T=2.5\times 10^9$K, respectively. Note that the $^{90}$Zr($\gamma$,$\alpha$) and $^{94}$Mo($\gamma$,p) channels are negligible with respect to the other channels mentioned above.

\begin{figure}[h]
\includegraphics[scale=0.4,keepaspectratio=true]{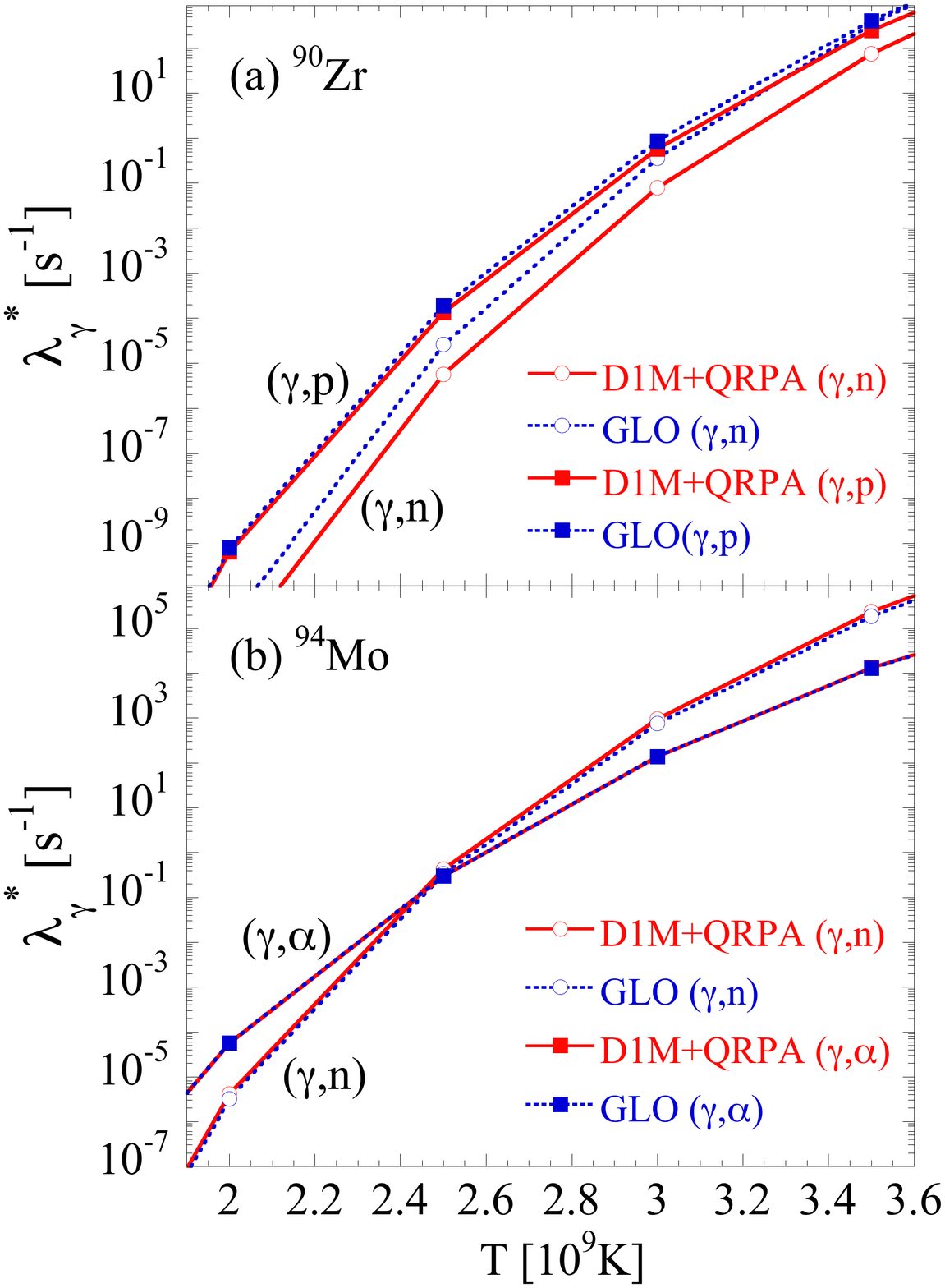}
\caption[Fig. 10]{
(Color online) (a) $^{90}$Zr($\gamma$,n)$^{89}$Zr (circles) and $^{90}$Zr($\gamma$,p)$^{89}$Y (squares) stellar reaction rates, as a function of the temperature, obtained with the D1M$+$QRPA (red solid lines) or the GLO (blue dotted lines) $\gamma$SF shown in Fig. 9. (b) Same for $^{94}$Mo($\gamma$,n)$^{93}$Mo and $^{94}$Mo($\gamma$,$\alpha$)$^{90}$Zr, respectively.}
\label{Fig10}
\end{figure}  

In this temperature range, we obtain a $^{94}$Mo($\gamma$,n)$^{93}$Mo stellar reaction rate with the D1M$+$QRPA dipole strength about 40\% higher than with the GLO model, whereas for $^{90}$Zr($\gamma$,n)$^{89}$Zr, a factor of 5 lower is obtained with D1M$+$QRPA adjusted to the present experimental data. This latter case shows how sensitive the stellar photoneutron reaction rate of astrophysical interest can be to experimental data in the vicinity of the neutron emission threshold, but also that even if the $\gamma$SF, hence the total photoabsorption cross sections, differ only by less than a factor of 2 (Figs.~\ref{Fig9}-\ref{Fig10}), the partial ($\gamma,n$) rates, strongly dominated by another emission channel, may differ by as much as a factor of 5. In the $^{94}$Mo case, the ($\gamma$,$\alpha$) rates obtained with both $\gamma$SF models are quite similar, and larger differences are again found on the photoneutron channel, the D1M+QRPA predictions giving this time higher rates.\newline
\indent  For a typical temperature of 2.5 GK for the $p$-process nucleosynthesis the ($\gamma$,n) reaction rates on the thermalized $^{90}$Zr and $^{94}$Mo are estimated with the D1M+QRPA dipole strength to be about 480 times and 200 times, respectively, larger than the rates on $^{90}$Zr and $^{94}$Mo in their ground state. Clearly, at these temperatures, transitions from and to the ground state (which are measured in the laboratory) contribute only to a small fraction of the stellar cross sections \cite{Raus11}. However, they are relevant for constraining HF statistical model parameters, such as the $\gamma$SF, as demonstrated with the present results especially in the case of the $^{90}$Zr($\gamma$,n) reaction.

\section{Conclusions}

\indent High-precision measurements of the photoneutron reaction cross sections on the nuclei $^{94}$Mo and $^{90}$Zr have been conducted from the respective neutron emission thresholds up to 13.5 MeV. Beams of high intensity quasi-monochromatic $\gamma$-rays from laser Compton scattering at the HI$\gamma$S facility were used. The new experimental cross sections were accurately measured near the neutron emission threshold which is where the photoneutron reaction cross sections are very small but astrophysically relevant. In order to constrain the $\gamma$SF in the A $\approx$ 90 mass region, the measured cross sections were compared with predictions of Hauser-Feshbach statistical model calculations using two different dipole $\gamma$SF models $-$ GLO model and HFB$+$QRPA model.\newline
\indent  For the $^{94}$Mo($\gamma$,n)$^{93}$Mo reaction the resulting stellar reaction rate, calculated with the D1M$+$QRPA dipole strength, was found to be about 40\% higher than the stellar reaction rate calculated with the GLO model. In contrast, for the $^{90}$Zr($\gamma$,n)$^{89}$Zr reaction, the stellar reaction rate calculated with the D1M$+$QRPA dipole strength and adjusted to the new experimental data was found to be a factor of 5 lower than the stellar reaction rate calculated with the GLO model. Hence, the present results show how sensitive the stellar photoneutron reaction rates of astrophysical interest are to experimental data in the vicinity of the neutron emission threshold.\newline
\indent Considering the very large number of nuclear reactions involved in the production of a single $p$-nucleus, measurements of photodisintegration reaction cross sections at astrophysically relevant energies on nuclei located as close as possible to the $p$-process path will put the nucleosynthesis calculations on a firmer ground. Therefore the present results on $^{94}$Mo, one of the most abundant of the $p$-nuclei and currently underproduced in all of the existing astrophysical models, and on $^{90}$Zr, a neutron magic nucleus known until recently as a genuine $s$-process nuclide, may have a significant impact on the efforts of understanding the $p$-process nucleosynthesis.\newline

\begin{acknowledgments}

\indent We thank C. Travaglio for her support and stimulating discussions in the early stage of this research. We would like to acknowledge the contributions of W. R. Zimmerman, M. Bhike, W. Tornow, and M. McCleskey during the collection of the data, as well as the staff at the HI$\gamma$S facility for their help during the experimental setup and for the production of the $\gamma$-ray beam. Additionally, we would like to thank R. Schwengner for providing the $^{94}$Mo and $^{90}$Zr targets, M. Starnes for machining the Al collimator, J. A. Gallant for preparing the experimental figures, T. M.-R. Chu and J. E. Mayer for assisting in the data reduction.\newline
\indent  Figs. 4$-$6 were created using R$:$ A language and environment for statistical computing. R Foundation for Statistical Computing, Vienna, Austria. R Core Team (2017)$;$ URL https://www.R-project.org/ while Fig. 3 was created using the software processing; URL https://www.processing.org/.\newline
\indent AB acknowledges support in part by the Research Corporation for Science Advancement, Grant No. 22662, Office of Sponsored Programs at James Madison University (JMU), Grant No. 100672, and by the Department of Physics and Astronomy at JMU.\newline
\indent JAS acknowledges that this work was performed under the auspices of the U.S. Department of Energy by Lawrence Livermore National Laboratory under Contract No. DE-AC52-07NA27344.\newline
\indent HJK acknowledges support of the U.S. Department of Energy, Office of Science, Office of Nuclear Physics through grants DE-FG02-97ER41033 and DE-FG02-97ER41041.\newline
\indent SG acknowledges the support of the FRS-FNRS. The theoretical work was performed within the IAEA CRP on ``Updating the Photonuclear data Library and Generating a Reference Database for Photon Strength Functions" (F41032). \newline
\indent Last but not least, we highly appreciated the thorough review of the anonymous referee.

\end{acknowledgments}

\bibliography{MyHIGSpaper_resub_final}

\begin{thebibliography}{52}%
\makeatletter
\providecommand \@ifxundefined [1]{%
 \@ifx{#1\undefined}
}%
\providecommand \@ifnum [1]{%
 \ifnum #1\expandafter \@firstoftwo
 \else \expandafter \@secondoftwo
 \fi
}%
\providecommand \@ifx [1]{%
 \ifx #1\expandafter \@firstoftwo
 \else \expandafter \@secondoftwo
 \fi
}%
\providecommand \natexlab [1]{#1}%
\providecommand \enquote  [1]{``#1''}%
\providecommand \bibnamefont  [1]{#1}%
\providecommand \bibfnamefont [1]{#1}%
\providecommand \citenamefont [1]{#1}%
\providecommand \href@noop [0]{\@secondoftwo}%
\providecommand \href [0]{\begingroup \@sanitize@url \@href}%
\providecommand \@href[1]{\@@startlink{#1}\@@href}%
\providecommand \@@href[1]{\endgroup#1\@@endlink}%
\providecommand \@sanitize@url [0]{\catcode `\\12\catcode `\$12\catcode
  `\&12\catcode `\#12\catcode `\^12\catcode `\_12\catcode `\%12\relax}%
\providecommand \@@startlink[1]{}%
\providecommand \@@endlink[0]{}%
\providecommand \url  [0]{\begingroup\@sanitize@url \@url }%
\providecommand \@url [1]{\endgroup\@href {#1}{\urlprefix }}%
\providecommand \urlprefix  [0]{URL }%
\providecommand \Eprint [0]{\href }%
\providecommand \doibase [0]{http://dx.doi.org/}%
\providecommand \selectlanguage [0]{\@gobble}%
\providecommand \bibinfo  [0]{\@secondoftwo}%
\providecommand \bibfield  [0]{\@secondoftwo}%
\providecommand \translation [1]{[#1]}%
\providecommand \BibitemOpen [0]{}%
\providecommand \bibitemStop [0]{}%
\providecommand \bibitemNoStop [0]{.\EOS\space}%
\providecommand \EOS [0]{\spacefactor3000\relax}%
\providecommand \BibitemShut  [1]{\csname bibitem#1\endcsname}%
\let\auto@bib@innerbib\@empty
\bibitem [{\citenamefont {Burbidge}\ \emph {et~al.}(1957)\citenamefont
  {Burbidge}, \citenamefont {Burbidge}, \citenamefont {Fowler},\ and\
  \citenamefont {Hoyle}}]{Burb57}%
  \BibitemOpen
  \bibfield  {author} {\bibinfo {author} {\bibfnamefont {M.~E.~M.}\
  \bibnamefont {Burbidge}}, \bibinfo {author} {\bibfnamefont {G.~R.}\
  \bibnamefont {Burbidge}}, \bibinfo {author} {\bibfnamefont {W.~A.}\
  \bibnamefont {Fowler}}, \ and\ \bibinfo {author} {\bibfnamefont
  {F.}~\bibnamefont {Hoyle}},\ }\href@noop {} {\bibfield  {journal} {\bibinfo
  {journal} {Rev. Mod. Phys.}\ }\textbf {\bibinfo {volume} {29}},\ \bibinfo
  {pages} {547} (\bibinfo {year} {1957})}\BibitemShut {NoStop}%
\bibitem [{\citenamefont {Arnould}(1976)}]{Arno76}%
  \BibitemOpen
  \bibfield  {author} {\bibinfo {author} {\bibfnamefont {M.}~\bibnamefont
  {Arnould}},\ }\href@noop {} {\bibfield  {journal} {\bibinfo  {journal}
  {Astron. Astrophys.}\ }\textbf {\bibinfo {volume} {46}},\ \bibinfo {pages}
  {17} (\bibinfo {year} {1976})}\BibitemShut {NoStop}%
\bibitem [{\citenamefont {Woosley}\ and\ \citenamefont
  {Howard}(1978)}]{Woos78}%
  \BibitemOpen
  \bibfield  {author} {\bibinfo {author} {\bibfnamefont {S.~E.}\ \bibnamefont
  {Woosley}}\ and\ \bibinfo {author} {\bibfnamefont {W.~M.}\ \bibnamefont
  {Howard}},\ }\href@noop {} {\bibfield  {journal} {\bibinfo  {journal} {The
  Astrophys. J. Suppl. S.}\ }\textbf {\bibinfo {volume} {36}},\ \bibinfo
  {pages} {285} (\bibinfo {year} {1978})}\BibitemShut {NoStop}%
\bibitem [{\citenamefont {Lambert}(1992)}]{Lamb92}%
  \BibitemOpen
  \bibfield  {author} {\bibinfo {author} {\bibfnamefont {D.~L.}\ \bibnamefont
  {Lambert}},\ }\href@noop {} {\bibfield  {journal} {\bibinfo  {journal}
  {Astron. Astrophys. Rev.}\ }\textbf {\bibinfo {volume} {3}},\ \bibinfo
  {pages} {201} (\bibinfo {year} {1992})}\BibitemShut {NoStop}%
\bibitem [{\citenamefont {Arnould}\ and\ \citenamefont
  {Goriely}(2003)}]{Arno03}%
  \BibitemOpen
  \bibfield  {author} {\bibinfo {author} {\bibfnamefont {M.}~\bibnamefont
  {Arnould}}\ and\ \bibinfo {author} {\bibfnamefont {S.}~\bibnamefont
  {Goriely}},\ }\href@noop {} {\bibfield  {journal} {\bibinfo  {journal} {Phys.
  Rep.}\ }\textbf {\bibinfo {volume} {384}},\ \bibinfo {pages} {1} (\bibinfo
  {year} {2003})}\BibitemShut {NoStop}%
\bibitem [{\citenamefont {Rayet}\ \emph {et~al.}(1995)\citenamefont {Rayet},
  \citenamefont {Arnould}, \citenamefont {Hashimoto}, \citenamefont
  {Prantzos},\ and\ \citenamefont {Nomoto}}]{Raye95}%
  \BibitemOpen
  \bibfield  {author} {\bibinfo {author} {\bibfnamefont {M.}~\bibnamefont
  {Rayet}}, \bibinfo {author} {\bibfnamefont {M.}~\bibnamefont {Arnould}},
  \bibinfo {author} {\bibfnamefont {M.}~\bibnamefont {Hashimoto}}, \bibinfo
  {author} {\bibfnamefont {N.}~\bibnamefont {Prantzos}}, \ and\ \bibinfo
  {author} {\bibfnamefont {K.}~\bibnamefont {Nomoto}},\ }\href@noop {}
  {\bibfield  {journal} {\bibinfo  {journal} {Astron. Astrophys.}\ }\textbf
  {\bibinfo {volume} {298}},\ \bibinfo {pages} {517} (\bibinfo {year}
  {1995})}\BibitemShut {NoStop}%
\bibitem [{\citenamefont {Rauscher}\ and\ \citenamefont
  {Thielemann}(2000)}]{Raus00}%
  \BibitemOpen
  \bibfield  {author} {\bibinfo {author} {\bibfnamefont {T.}~\bibnamefont
  {Rauscher}}\ and\ \bibinfo {author} {\bibfnamefont {F.-K.}\ \bibnamefont
  {Thielemann}},\ }\href@noop {} {\bibfield  {journal} {\bibinfo  {journal}
  {At. Data Nucl. Data Tables}\ }\textbf {\bibinfo {volume} {75}},\ \bibinfo
  {pages} {1} (\bibinfo {year} {2000})}\BibitemShut {NoStop}%
\bibitem [{\citenamefont {Goriely}\ \emph {et~al.}(2002)\citenamefont
  {Goriely}, \citenamefont {Jos\'e}, \citenamefont {Hernanz}, \citenamefont
  {Rayet},\ and\ \citenamefont {Arnould}}]{Gori02}%
  \BibitemOpen
  \bibfield  {author} {\bibinfo {author} {\bibfnamefont {S.}~\bibnamefont
  {Goriely}}, \bibinfo {author} {\bibfnamefont {J.}~\bibnamefont {Jos\'e}},
  \bibinfo {author} {\bibfnamefont {M.}~\bibnamefont {Hernanz}}, \bibinfo
  {author} {\bibfnamefont {M.}~\bibnamefont {Rayet}}, \ and\ \bibinfo {author}
  {\bibfnamefont {M.}~\bibnamefont {Arnould}},\ }\href@noop {} {\bibfield
  {journal} {\bibinfo  {journal} {Astron. Astrophys.}\ }\textbf {\bibinfo
  {volume} {383}},\ \bibinfo {pages} {L27} (\bibinfo {year}
  {2002})}\BibitemShut {NoStop}%
\bibitem [{\citenamefont {Fr\"ohlich}\ \emph {et~al.}(2006)\citenamefont
  {Fr\"ohlich}, \citenamefont {Mart\'inez-Pinedo}, \citenamefont
  {Liebend\"orfer}, \citenamefont {Thielemann}, \citenamefont {Bravo},
  \citenamefont {Hix}, \citenamefont {Langanke},\ and\ \citenamefont
  {Zinner}}]{Froh06}%
  \BibitemOpen
  \bibfield  {author} {\bibinfo {author} {\bibfnamefont {C.}~\bibnamefont
  {Fr\"ohlich}}, \bibinfo {author} {\bibfnamefont {G.}~\bibnamefont
  {Mart\'inez-Pinedo}}, \bibinfo {author} {\bibfnamefont {M.}~\bibnamefont
  {Liebend\"orfer}}, \bibinfo {author} {\bibfnamefont {F.-K.}\ \bibnamefont
  {Thielemann}}, \bibinfo {author} {\bibfnamefont {E.}~\bibnamefont {Bravo}},
  \bibinfo {author} {\bibfnamefont {W.~R.}\ \bibnamefont {Hix}}, \bibinfo
  {author} {\bibfnamefont {K.}~\bibnamefont {Langanke}}, \ and\ \bibinfo
  {author} {\bibfnamefont {N.~T.}\ \bibnamefont {Zinner}},\ }\href@noop {}
  {\bibfield  {journal} {\bibinfo  {journal} {Phys. Rev. Lett.}\ }\textbf
  {\bibinfo {volume} {96}},\ \bibinfo {pages} {142502} (\bibinfo {year}
  {2006})}\BibitemShut {NoStop}%
\bibitem [{\citenamefont {Travaglio}\ \emph {et~al.}(2011)\citenamefont
  {Travaglio}, \citenamefont {R\"opke}, \citenamefont {Gallino},\ and\
  \citenamefont {Hillebrandt}}]{Trav11}%
  \BibitemOpen
  \bibfield  {author} {\bibinfo {author} {\bibfnamefont {C.}~\bibnamefont
  {Travaglio}}, \bibinfo {author} {\bibfnamefont {F.~K.}\ \bibnamefont
  {R\"opke}}, \bibinfo {author} {\bibfnamefont {R.}~\bibnamefont {Gallino}}, \
  and\ \bibinfo {author} {\bibfnamefont {W.}~\bibnamefont {Hillebrandt}},\
  }\href@noop {} {\bibfield  {journal} {\bibinfo  {journal} {Astroph. J.}\
  }\textbf {\bibinfo {volume} {739}},\ \bibinfo {pages} {93} (\bibinfo {year}
  {2011})}\BibitemShut {NoStop}%
\bibitem [{\citenamefont {Goriely}\ and\ \citenamefont {Khan}(2002)}]{Gori02b}%
  \BibitemOpen
  \bibfield  {author} {\bibinfo {author} {\bibfnamefont {S.}~\bibnamefont
  {Goriely}}\ and\ \bibinfo {author} {\bibfnamefont {E.}~\bibnamefont {Khan}},\
  }\href@noop {} {\bibfield  {journal} {\bibinfo  {journal} {Nucl. Phys.}\
  }\textbf {\bibinfo {volume} {A706}},\ \bibinfo {pages} {217} (\bibinfo {year}
  {2002})}\BibitemShut {NoStop}%
\bibitem [{\citenamefont {Tsoneva}\ \emph {et~al.}(2015)\citenamefont
  {Tsoneva}, \citenamefont {Goriely}, \citenamefont {Lenske},\ and\
  \citenamefont {Schwengner}}]{Tson15}%
  \BibitemOpen
  \bibfield  {author} {\bibinfo {author} {\bibfnamefont {N.}~\bibnamefont
  {Tsoneva}}, \bibinfo {author} {\bibfnamefont {S.}~\bibnamefont {Goriely}},
  \bibinfo {author} {\bibfnamefont {H.}~\bibnamefont {Lenske}}, \ and\ \bibinfo
  {author} {\bibfnamefont {R.}~\bibnamefont {Schwengner}},\ }\href@noop {}
  {\bibfield  {journal} {\bibinfo  {journal} {Phys. Rev. C}\ }\textbf {\bibinfo
  {volume} {91}},\ \bibinfo {pages} {044318} (\bibinfo {year}
  {2015})}\BibitemShut {NoStop}%
\bibitem [{\citenamefont {Martini}\ \emph {et~al.}(2016)\citenamefont
  {Martini}, \citenamefont {P\'eru}, \citenamefont {Hilaire}, \citenamefont
  {Goriely},\ and\ \citenamefont {Lechaftois}}]{Mart16}%
  \BibitemOpen
  \bibfield  {author} {\bibinfo {author} {\bibfnamefont {M.}~\bibnamefont
  {Martini}}, \bibinfo {author} {\bibfnamefont {S.}~\bibnamefont {P\'eru}},
  \bibinfo {author} {\bibfnamefont {S.}~\bibnamefont {Hilaire}}, \bibinfo
  {author} {\bibfnamefont {S.}~\bibnamefont {Goriely}}, \ and\ \bibinfo
  {author} {\bibfnamefont {F.}~\bibnamefont {Lechaftois}},\ }\href@noop {}
  {\bibfield  {journal} {\bibinfo  {journal} {Phys. Rev. C}\ }\textbf {\bibinfo
  {volume} {94}},\ \bibinfo {pages} {014304} (\bibinfo {year}
  {2016})}\BibitemShut {NoStop}%
\bibitem [{\citenamefont {Goriely}\ \emph {et~al.}(2016)\citenamefont
  {Goriely}, \citenamefont {Hilaire}, \citenamefont {P\'eru}, \citenamefont
  {Martini}, \citenamefont {Deloncle},\ and\ \citenamefont
  {Lechaftois}}]{Gori16}%
  \BibitemOpen
  \bibfield  {author} {\bibinfo {author} {\bibfnamefont {S.}~\bibnamefont
  {Goriely}}, \bibinfo {author} {\bibfnamefont {S.}~\bibnamefont {Hilaire}},
  \bibinfo {author} {\bibfnamefont {S.}~\bibnamefont {P\'eru}}, \bibinfo
  {author} {\bibfnamefont {M.}~\bibnamefont {Martini}}, \bibinfo {author}
  {\bibfnamefont {I.}~\bibnamefont {Deloncle}}, \ and\ \bibinfo {author}
  {\bibfnamefont {F.}~\bibnamefont {Lechaftois}},\ }\href@noop {} {\bibfield
  {journal} {\bibinfo  {journal} {Phys. Rev. C}\ }\textbf {\bibinfo {volume}
  {94}},\ \bibinfo {pages} {044306} (\bibinfo {year} {2016})}\BibitemShut
  {NoStop}%
\bibitem [{\citenamefont {Goriely}\ \emph {et~al.}(2018)\citenamefont
  {Goriely}, \citenamefont {Hilaire}, \citenamefont {P\'eru},\ and\
  \citenamefont {Sieja}}]{Gori18}%
  \BibitemOpen
  \bibfield  {author} {\bibinfo {author} {\bibfnamefont {S.}~\bibnamefont
  {Goriely}}, \bibinfo {author} {\bibfnamefont {S.}~\bibnamefont {Hilaire}},
  \bibinfo {author} {\bibfnamefont {S.}~\bibnamefont {P\'eru}}, \ and\ \bibinfo
  {author} {\bibfnamefont {K.}~\bibnamefont {Sieja}},\ }\href@noop {}
  {\bibfield  {journal} {\bibinfo  {journal} {Phys. Rev. C}\ }\textbf {\bibinfo
  {volume} {98}},\ \bibinfo {pages} {014327} (\bibinfo {year}
  {2018})}\BibitemShut {NoStop}%
\bibitem [{\citenamefont {Utsunomiya}\ \emph {et~al.}(2009)\citenamefont
  {Utsunomiya}, \citenamefont {Goriely}, \citenamefont {Kamata}, \citenamefont
  {Kondo}, \citenamefont {Itoh}, \citenamefont {Akimune}, \citenamefont
  {Yamagata}, \citenamefont {Toyokawa}, \citenamefont {Lui}, \citenamefont
  {Hilaire},\ and\ \citenamefont {Koning}}]{Utsu09}%
  \BibitemOpen
  \bibfield  {author} {\bibinfo {author} {\bibfnamefont {H.}~\bibnamefont
  {Utsunomiya}}, \bibinfo {author} {\bibfnamefont {S.}~\bibnamefont {Goriely}},
  \bibinfo {author} {\bibfnamefont {M.}~\bibnamefont {Kamata}}, \bibinfo
  {author} {\bibfnamefont {T.}~\bibnamefont {Kondo}}, \bibinfo {author}
  {\bibfnamefont {O.}~\bibnamefont {Itoh}}, \bibinfo {author} {\bibfnamefont
  {H.}~\bibnamefont {Akimune}}, \bibinfo {author} {\bibfnamefont
  {T.}~\bibnamefont {Yamagata}}, \bibinfo {author} {\bibfnamefont
  {H.}~\bibnamefont {Toyokawa}}, \bibinfo {author} {\bibfnamefont {Y.-W.}\
  \bibnamefont {Lui}}, \bibinfo {author} {\bibfnamefont {S.}~\bibnamefont
  {Hilaire}}, \ and\ \bibinfo {author} {\bibfnamefont {A.~J.}\ \bibnamefont
  {Koning}},\ }\href@noop {} {\bibfield  {journal} {\bibinfo  {journal} {Phys.
  Rev. C}\ }\textbf {\bibinfo {volume} {80}},\ \bibinfo {pages} {055806}
  (\bibinfo {year} {2009})}\BibitemShut {NoStop}%
\bibitem [{\citenamefont {Tonchev}\ \emph {et~al.}(2010)\citenamefont
  {Tonchev}, \citenamefont {Hammond}, \citenamefont {Howell}, \citenamefont
  {Huibregtse}, \citenamefont {Hutcheson}, \citenamefont {Kelley},
  \citenamefont {Kwan}, \citenamefont {Raut}, \citenamefont {Rusev},
  \citenamefont {Tornow}, \citenamefont {Kawano}, \citenamefont {Vieira},\ and\
  \citenamefont {Wilhelmy}}]{Tonc10}%
  \BibitemOpen
  \bibfield  {author} {\bibinfo {author} {\bibfnamefont {A.~P.}\ \bibnamefont
  {Tonchev}}, \bibinfo {author} {\bibfnamefont {S.~L.}\ \bibnamefont
  {Hammond}}, \bibinfo {author} {\bibfnamefont {C.~R.}\ \bibnamefont {Howell}},
  \bibinfo {author} {\bibfnamefont {C.}~\bibnamefont {Huibregtse}}, \bibinfo
  {author} {\bibfnamefont {A.}~\bibnamefont {Hutcheson}}, \bibinfo {author}
  {\bibfnamefont {J.~H.}\ \bibnamefont {Kelley}}, \bibinfo {author}
  {\bibfnamefont {E.}~\bibnamefont {Kwan}}, \bibinfo {author} {\bibfnamefont
  {R.}~\bibnamefont {Raut}}, \bibinfo {author} {\bibfnamefont {G.}~\bibnamefont
  {Rusev}}, \bibinfo {author} {\bibfnamefont {W.}~\bibnamefont {Tornow}},
  \bibinfo {author} {\bibfnamefont {T.}~\bibnamefont {Kawano}}, \bibinfo
  {author} {\bibfnamefont {D.~J.}\ \bibnamefont {Vieira}}, \ and\ \bibinfo
  {author} {\bibfnamefont {J.~B.}\ \bibnamefont {Wilhelmy}},\ }\href@noop {}
  {\bibfield  {journal} {\bibinfo  {journal} {Phys. Rev. C}\ }\textbf {\bibinfo
  {volume} {82}},\ \bibinfo {pages} {054620} (\bibinfo {year}
  {2010})}\BibitemShut {NoStop}%
\bibitem [{\citenamefont {Utsunomiya}\ \emph {et~al.}(2013)\citenamefont
  {Utsunomiya}, \citenamefont {Goriely}, \citenamefont {Kondo}, \citenamefont
  {Iwamoto}, \citenamefont {Akimune}, \citenamefont {Yamagata}, \citenamefont
  {Toyokawa}, \citenamefont {Harada}, \citenamefont {Kitatani}, \citenamefont
  {Lui}, \citenamefont {Larsen}, \citenamefont {Guttormsen}, \citenamefont
  {Koehler}, \citenamefont {Hilaire}, \citenamefont {P\'eru}, \citenamefont
  {Martini},\ and\ \citenamefont {Koning}}]{Utsu13}%
  \BibitemOpen
  \bibfield  {author} {\bibinfo {author} {\bibfnamefont {H.}~\bibnamefont
  {Utsunomiya}}, \bibinfo {author} {\bibfnamefont {S.}~\bibnamefont {Goriely}},
  \bibinfo {author} {\bibfnamefont {T.}~\bibnamefont {Kondo}}, \bibinfo
  {author} {\bibfnamefont {C.}~\bibnamefont {Iwamoto}}, \bibinfo {author}
  {\bibfnamefont {H.}~\bibnamefont {Akimune}}, \bibinfo {author} {\bibfnamefont
  {T.}~\bibnamefont {Yamagata}}, \bibinfo {author} {\bibfnamefont
  {H.}~\bibnamefont {Toyokawa}}, \bibinfo {author} {\bibfnamefont
  {H.}~\bibnamefont {Harada}}, \bibinfo {author} {\bibfnamefont
  {F.}~\bibnamefont {Kitatani}}, \bibinfo {author} {\bibfnamefont {Y.-W.}\
  \bibnamefont {Lui}}, \bibinfo {author} {\bibfnamefont {A.~C.}\ \bibnamefont
  {Larsen}}, \bibinfo {author} {\bibfnamefont {M.}~\bibnamefont {Guttormsen}},
  \bibinfo {author} {\bibfnamefont {P.~E.}\ \bibnamefont {Koehler}}, \bibinfo
  {author} {\bibfnamefont {S.}~\bibnamefont {Hilaire}}, \bibinfo {author}
  {\bibfnamefont {S.}~\bibnamefont {P\'eru}}, \bibinfo {author} {\bibfnamefont
  {M.}~\bibnamefont {Martini}}, \ and\ \bibinfo {author} {\bibfnamefont
  {A.~J.}\ \bibnamefont {Koning}},\ }\href@noop {} {\bibfield  {journal}
  {\bibinfo  {journal} {Phys. Rev. C}\ }\textbf {\bibinfo {volume} {88}},\
  \bibinfo {pages} {015805} (\bibinfo {year} {2013})}\BibitemShut {NoStop}%
\bibitem [{\citenamefont {Raut}\ \emph {et~al.}(2013)\citenamefont {Raut},
  \citenamefont {Tonchev}, \citenamefont {Rusev}, \citenamefont {Tornow},
  \citenamefont {Iliadis}, \citenamefont {Lugaro}, \citenamefont {Buntain},
  \citenamefont {Goriely}, \citenamefont {Kelley}, \citenamefont {Schwengner},
  \citenamefont {Banu},\ and\ \citenamefont {Tsoneva}}]{Raut13}%
  \BibitemOpen
  \bibfield  {author} {\bibinfo {author} {\bibfnamefont {R.}~\bibnamefont
  {Raut}}, \bibinfo {author} {\bibfnamefont {A.~P.}\ \bibnamefont {Tonchev}},
  \bibinfo {author} {\bibfnamefont {G.}~\bibnamefont {Rusev}}, \bibinfo
  {author} {\bibfnamefont {W.}~\bibnamefont {Tornow}}, \bibinfo {author}
  {\bibfnamefont {C.}~\bibnamefont {Iliadis}}, \bibinfo {author} {\bibfnamefont
  {M.}~\bibnamefont {Lugaro}}, \bibinfo {author} {\bibfnamefont
  {J.}~\bibnamefont {Buntain}}, \bibinfo {author} {\bibfnamefont
  {S.}~\bibnamefont {Goriely}}, \bibinfo {author} {\bibfnamefont {J.~H.}\
  \bibnamefont {Kelley}}, \bibinfo {author} {\bibfnamefont {R.}~\bibnamefont
  {Schwengner}}, \bibinfo {author} {\bibfnamefont {A.}~\bibnamefont {Banu}}, \
  and\ \bibinfo {author} {\bibfnamefont {N.}~\bibnamefont {Tsoneva}},\
  }\href@noop {} {\bibfield  {journal} {\bibinfo  {journal} {Phys. Rev. Lett.}\
  }\textbf {\bibinfo {volume} {111}},\ \bibinfo {pages} {112501} (\bibinfo
  {year} {2013})}\BibitemShut {NoStop}%
\bibitem [{\citenamefont {Filipescu}\ \emph {et~al.}(2014)\citenamefont
  {Filipescu}, \citenamefont {Gheorghe}, \citenamefont {Utsunomiya},
  \citenamefont {Goriely}, \citenamefont {Renstrom}, \citenamefont {Nyhus},
  \citenamefont {Tesileanu}, \citenamefont {Glodariu}, \citenamefont {Shima},
  \citenamefont {Takahisa}, \citenamefont {Miyamoto}, \citenamefont {Lui},
  \citenamefont {Hilaire}, \citenamefont {P\'eru}, \citenamefont {Martini},\
  and\ \citenamefont {Koning}}]{Fili14}%
  \BibitemOpen
  \bibfield  {author} {\bibinfo {author} {\bibfnamefont {D.~M.}\ \bibnamefont
  {Filipescu}}, \bibinfo {author} {\bibfnamefont {I.}~\bibnamefont {Gheorghe}},
  \bibinfo {author} {\bibfnamefont {H.}~\bibnamefont {Utsunomiya}}, \bibinfo
  {author} {\bibfnamefont {S.}~\bibnamefont {Goriely}}, \bibinfo {author}
  {\bibfnamefont {T.}~\bibnamefont {Renstrom}}, \bibinfo {author}
  {\bibfnamefont {H.-T.}\ \bibnamefont {Nyhus}}, \bibinfo {author}
  {\bibfnamefont {O.}~\bibnamefont {Tesileanu}}, \bibinfo {author}
  {\bibfnamefont {T.}~\bibnamefont {Glodariu}}, \bibinfo {author}
  {\bibfnamefont {T.}~\bibnamefont {Shima}}, \bibinfo {author} {\bibfnamefont
  {K.}~\bibnamefont {Takahisa}}, \bibinfo {author} {\bibfnamefont
  {S.}~\bibnamefont {Miyamoto}}, \bibinfo {author} {\bibfnamefont {Y.-W.}\
  \bibnamefont {Lui}}, \bibinfo {author} {\bibfnamefont {S.}~\bibnamefont
  {Hilaire}}, \bibinfo {author} {\bibfnamefont {S.}~\bibnamefont {P\'eru}},
  \bibinfo {author} {\bibfnamefont {M.}~\bibnamefont {Martini}}, \ and\
  \bibinfo {author} {\bibfnamefont {A.~J.}\ \bibnamefont {Koning}},\
  }\href@noop {} {\bibfield  {journal} {\bibinfo  {journal} {Phys. Rev. C}\
  }\textbf {\bibinfo {volume} {90}},\ \bibinfo {pages} {064616} (\bibinfo
  {year} {2014})}\BibitemShut {NoStop}%
\bibitem [{\citenamefont {Sauerwein}\ \emph {et~al.}(2014)\citenamefont
  {Sauerwein}, \citenamefont {Sonnabend}, \citenamefont {Fritzsche},
  \citenamefont {Glorius}, \citenamefont {Kwan}, \citenamefont {Pietralla},
  \citenamefont {Romig}, \citenamefont {Rusev}, \citenamefont {Savran},
  \citenamefont {Schnorrenberger}, \citenamefont {Tonchev}, \citenamefont
  {Tornow},\ and\ \citenamefont {Weller}}]{Saue14}%
  \BibitemOpen
  \bibfield  {author} {\bibinfo {author} {\bibfnamefont {A.}~\bibnamefont
  {Sauerwein}}, \bibinfo {author} {\bibfnamefont {K.}~\bibnamefont
  {Sonnabend}}, \bibinfo {author} {\bibfnamefont {M.}~\bibnamefont
  {Fritzsche}}, \bibinfo {author} {\bibfnamefont {J.}~\bibnamefont {Glorius}},
  \bibinfo {author} {\bibfnamefont {E.}~\bibnamefont {Kwan}}, \bibinfo {author}
  {\bibfnamefont {N.}~\bibnamefont {Pietralla}}, \bibinfo {author}
  {\bibfnamefont {C.}~\bibnamefont {Romig}}, \bibinfo {author} {\bibfnamefont
  {G.}~\bibnamefont {Rusev}}, \bibinfo {author} {\bibfnamefont
  {D.}~\bibnamefont {Savran}}, \bibinfo {author} {\bibfnamefont
  {L.}~\bibnamefont {Schnorrenberger}}, \bibinfo {author} {\bibfnamefont
  {A.~P.}\ \bibnamefont {Tonchev}}, \bibinfo {author} {\bibfnamefont
  {W.}~\bibnamefont {Tornow}}, \ and\ \bibinfo {author} {\bibfnamefont {H.~R.}\
  \bibnamefont {Weller}},\ }\href@noop {} {\bibfield  {journal} {\bibinfo
  {journal} {Phys. Rev. C}\ }\textbf {\bibinfo {volume} {89}},\ \bibinfo
  {pages} {035803} (\bibinfo {year} {2014})}\BibitemShut {NoStop}%
\bibitem [{\citenamefont {Nyhus}\ \emph {et~al.}(2015)\citenamefont {Nyhus},
  \citenamefont {Renstrom}, \citenamefont {Utsunomiya}, \citenamefont
  {Goriely}, \citenamefont {Filipescu}, \citenamefont {Gheorghe}, \citenamefont
  {Tesileanu}, \citenamefont {Glodariu}, \citenamefont {Shima}, \citenamefont
  {Takahisa}, \citenamefont {Miyamoto}, \citenamefont {Lui}, \citenamefont
  {Hilaire}, \citenamefont {Peru}, \citenamefont {Martini}, \citenamefont
  {Siess},\ and\ \citenamefont {Koning}}]{Nyhu15}%
  \BibitemOpen
  \bibfield  {author} {\bibinfo {author} {\bibfnamefont {H.-T.}\ \bibnamefont
  {Nyhus}}, \bibinfo {author} {\bibfnamefont {T.}~\bibnamefont {Renstrom}},
  \bibinfo {author} {\bibfnamefont {H.}~\bibnamefont {Utsunomiya}}, \bibinfo
  {author} {\bibfnamefont {S.}~\bibnamefont {Goriely}}, \bibinfo {author}
  {\bibfnamefont {D.~M.}\ \bibnamefont {Filipescu}}, \bibinfo {author}
  {\bibfnamefont {I.}~\bibnamefont {Gheorghe}}, \bibinfo {author}
  {\bibfnamefont {O.}~\bibnamefont {Tesileanu}}, \bibinfo {author}
  {\bibfnamefont {T.}~\bibnamefont {Glodariu}}, \bibinfo {author}
  {\bibfnamefont {T.}~\bibnamefont {Shima}}, \bibinfo {author} {\bibfnamefont
  {K.}~\bibnamefont {Takahisa}}, \bibinfo {author} {\bibfnamefont
  {S.}~\bibnamefont {Miyamoto}}, \bibinfo {author} {\bibfnamefont {Y.-W.}\
  \bibnamefont {Lui}}, \bibinfo {author} {\bibfnamefont {S.}~\bibnamefont
  {Hilaire}}, \bibinfo {author} {\bibfnamefont {S.}~\bibnamefont {Peru}},
  \bibinfo {author} {\bibfnamefont {M.}~\bibnamefont {Martini}}, \bibinfo
  {author} {\bibfnamefont {L.}~\bibnamefont {Siess}}, \ and\ \bibinfo {author}
  {\bibfnamefont {A.~J.}\ \bibnamefont {Koning}},\ }\href@noop {} {\bibfield
  {journal} {\bibinfo  {journal} {Phys. Rev. C}\ }\textbf {\bibinfo {volume}
  {91}},\ \bibinfo {pages} {015808} (\bibinfo {year} {2015})}\BibitemShut
  {NoStop}%
\bibitem [{\citenamefont {Renstr{\o}m}\ \emph {et~al.}(2018)\citenamefont
  {Renstr{\o}m}, \citenamefont {Utsunomiya}, \citenamefont {Nyhus},
  \citenamefont {Larsen}, \citenamefont {Guttormsen}, \citenamefont {Tveten},
  \citenamefont {Filipescu}, \citenamefont {Gheorghe}, \citenamefont {Goriely},
  \citenamefont {Hilaire}, \citenamefont {Lui}, \citenamefont {Midtbo},
  \citenamefont {P\'eru}, \citenamefont {Shima}, \citenamefont {Siem},\ and\
  \citenamefont {Tesileanu}}]{Rens18}%
  \BibitemOpen
  \bibfield  {author} {\bibinfo {author} {\bibfnamefont {T.}~\bibnamefont
  {Renstr{\o}m}}, \bibinfo {author} {\bibfnamefont {H.}~\bibnamefont
  {Utsunomiya}}, \bibinfo {author} {\bibfnamefont {H.~T.}\ \bibnamefont
  {Nyhus}}, \bibinfo {author} {\bibfnamefont {A.~C.}\ \bibnamefont {Larsen}},
  \bibinfo {author} {\bibfnamefont {M.}~\bibnamefont {Guttormsen}}, \bibinfo
  {author} {\bibfnamefont {G.~M.}\ \bibnamefont {Tveten}}, \bibinfo {author}
  {\bibfnamefont {D.~M.}\ \bibnamefont {Filipescu}}, \bibinfo {author}
  {\bibfnamefont {I.}~\bibnamefont {Gheorghe}}, \bibinfo {author}
  {\bibfnamefont {S.}~\bibnamefont {Goriely}}, \bibinfo {author} {\bibfnamefont
  {S.}~\bibnamefont {Hilaire}}, \bibinfo {author} {\bibfnamefont {Y.-W.}\
  \bibnamefont {Lui}}, \bibinfo {author} {\bibfnamefont {J.~E.}\ \bibnamefont
  {Midtbo}}, \bibinfo {author} {\bibfnamefont {S.}~\bibnamefont {P\'eru}},
  \bibinfo {author} {\bibfnamefont {T.}~\bibnamefont {Shima}}, \bibinfo
  {author} {\bibfnamefont {S.}~\bibnamefont {Siem}}, \ and\ \bibinfo {author}
  {\bibfnamefont {O.}~\bibnamefont {Tesileanu}},\ }\href@noop {} {\bibfield
  {journal} {\bibinfo  {journal} {Phys. Rev. C}\ }\textbf {\bibinfo {volume}
  {98}},\ \bibinfo {pages} {054310} (\bibinfo {year} {2018})}\BibitemShut
  {NoStop}%
\bibitem [{\citenamefont {Utsunomiya}\ \emph {et~al.}(2018)\citenamefont
  {Utsunomiya}, \citenamefont {Renstr{\o}m}, \citenamefont {Tveten},
  \citenamefont {Goriely}, \citenamefont {Katayama}, \citenamefont {Ari-izumi},
  \citenamefont {Takenaka}, \citenamefont {Symochko}, \citenamefont {Kheswa},
  \citenamefont {Ingeberg}, \citenamefont {Glodariu}, \citenamefont {Lui},
  \citenamefont {Miyamoto}, \citenamefont {Larsen}, \citenamefont {Midtbo},
  \citenamefont {Gorgen}, \citenamefont {Siem}, \citenamefont {Campo},
  \citenamefont {Guttormsen}, \citenamefont {Hilaire}, \citenamefont {P\'eru},\
  and\ \citenamefont {Koning}}]{Utsu18}%
  \BibitemOpen
  \bibfield  {author} {\bibinfo {author} {\bibfnamefont {H.}~\bibnamefont
  {Utsunomiya}}, \bibinfo {author} {\bibfnamefont {T.}~\bibnamefont
  {Renstr{\o}m}}, \bibinfo {author} {\bibfnamefont {G.~M.}\ \bibnamefont
  {Tveten}}, \bibinfo {author} {\bibfnamefont {S.}~\bibnamefont {Goriely}},
  \bibinfo {author} {\bibfnamefont {S.}~\bibnamefont {Katayama}}, \bibinfo
  {author} {\bibfnamefont {T.}~\bibnamefont {Ari-izumi}}, \bibinfo {author}
  {\bibfnamefont {D.}~\bibnamefont {Takenaka}}, \bibinfo {author}
  {\bibfnamefont {D.}~\bibnamefont {Symochko}}, \bibinfo {author}
  {\bibfnamefont {B.~V.}\ \bibnamefont {Kheswa}}, \bibinfo {author}
  {\bibfnamefont {V.~W.}\ \bibnamefont {Ingeberg}}, \bibinfo {author}
  {\bibfnamefont {T.}~\bibnamefont {Glodariu}}, \bibinfo {author}
  {\bibfnamefont {Y.-W.}\ \bibnamefont {Lui}}, \bibinfo {author} {\bibfnamefont
  {S.}~\bibnamefont {Miyamoto}}, \bibinfo {author} {\bibfnamefont {A.~C.}\
  \bibnamefont {Larsen}}, \bibinfo {author} {\bibfnamefont {J.~E.}\
  \bibnamefont {Midtbo}}, \bibinfo {author} {\bibfnamefont {A.}~\bibnamefont
  {Gorgen}}, \bibinfo {author} {\bibfnamefont {S.}~\bibnamefont {Siem}},
  \bibinfo {author} {\bibfnamefont {L.~C.}\ \bibnamefont {Campo}}, \bibinfo
  {author} {\bibfnamefont {M.}~\bibnamefont {Guttormsen}}, \bibinfo {author}
  {\bibfnamefont {S.}~\bibnamefont {Hilaire}}, \bibinfo {author} {\bibfnamefont
  {S.}~\bibnamefont {P\'eru}}, \ and\ \bibinfo {author} {\bibfnamefont {A.~J.}\
  \bibnamefont {Koning}},\ }\href@noop {} {\bibfield  {journal} {\bibinfo
  {journal} {Phys. Rev. C}\ }\textbf {\bibinfo {volume} {98}},\ \bibinfo
  {pages} {054619} (\bibinfo {year} {2018})}\BibitemShut {NoStop}%
\bibitem [{\citenamefont {Travaglio}\ \emph {et~al.}(2015)\citenamefont
  {Travaglio}, \citenamefont {Gallino}, \citenamefont {Rauscher}, \citenamefont
  {R\"opke},\ and\ \citenamefont {Hillebrandt}}]{Trav15}%
  \BibitemOpen
  \bibfield  {author} {\bibinfo {author} {\bibfnamefont {C.}~\bibnamefont
  {Travaglio}}, \bibinfo {author} {\bibfnamefont {R.}~\bibnamefont {Gallino}},
  \bibinfo {author} {\bibfnamefont {T.}~\bibnamefont {Rauscher}}, \bibinfo
  {author} {\bibfnamefont {F.~K.}\ \bibnamefont {R\"opke}}, \ and\ \bibinfo
  {author} {\bibfnamefont {W.}~\bibnamefont {Hillebrandt}},\ }\href@noop {}
  {\bibfield  {journal} {\bibinfo  {journal} {Astroph. J.}\ }\textbf {\bibinfo
  {volume} {799}},\ \bibinfo {pages} {54} (\bibinfo {year} {2015})}\BibitemShut
  {NoStop}%
\bibitem [{\citenamefont {Weller}\ \emph {et~al.}(2009)\citenamefont {Weller}
  \emph {et~al.}}]{Well09}%
  \BibitemOpen
  \bibfield  {author} {\bibinfo {author} {\bibfnamefont {H.~R.}\ \bibnamefont
  {Weller}} \emph {et~al.},\ }\href@noop {} {\bibfield  {journal} {\bibinfo
  {journal} {Prog. Part. Nucl. Phys.}\ }\textbf {\bibinfo {volume} {62}},\
  \bibinfo {pages} {257} (\bibinfo {year} {2009})}\BibitemShut {NoStop}%
\bibitem [{\citenamefont {Pywell}\ \emph {et~al.}(2009)\citenamefont {Pywell},
  \citenamefont {Mavrichi}, \citenamefont {Wurtz},\ and\ \citenamefont
  {Wilson}}]{Pywe09}%
  \BibitemOpen
  \bibfield  {author} {\bibinfo {author} {\bibfnamefont {R.~E.}\ \bibnamefont
  {Pywell}}, \bibinfo {author} {\bibfnamefont {O.}~\bibnamefont {Mavrichi}},
  \bibinfo {author} {\bibfnamefont {W.~A.}\ \bibnamefont {Wurtz}}, \ and\
  \bibinfo {author} {\bibfnamefont {R.}~\bibnamefont {Wilson}},\ }\href@noop {}
  {\bibfield  {journal} {\bibinfo  {journal} {Nucl.\ Instrum.\ Methods Phys.\
  Res. Sect. A}\ }\textbf {\bibinfo {volume} {606}},\ \bibinfo {pages} {517}
  (\bibinfo {year} {2009})}\BibitemShut {NoStop}%
\bibitem [{\citenamefont {GEANT4}(2003)}]{Gean03}%
  \BibitemOpen
  \bibfield  {author} {\bibinfo {author} {\bibnamefont {GEANT4}},\ }\href@noop
  {} {\bibfield  {journal} {\bibinfo  {journal} {Nucl.\ Instrum.\ Methods\ A}\
  }\textbf {\bibinfo {volume} {503}},\ \bibinfo {pages} {250} (\bibinfo {year}
  {2003})}\BibitemShut {NoStop}%
\bibitem [{\citenamefont {Birenbaum}\ \emph {et~al.}(1985)\citenamefont
  {Birenbaum}, \citenamefont {Kahane},\ and\ \citenamefont {Moreh}}]{Bire85}%
  \BibitemOpen
  \bibfield  {author} {\bibinfo {author} {\bibfnamefont {Y.}~\bibnamefont
  {Birenbaum}}, \bibinfo {author} {\bibfnamefont {S.}~\bibnamefont {Kahane}}, \
  and\ \bibinfo {author} {\bibfnamefont {R.}~\bibnamefont {Moreh}},\
  }\href@noop {} {\bibfield  {journal} {\bibinfo  {journal} {Phys. Rev. C}\
  }\textbf {\bibinfo {volume} {32}},\ \bibinfo {pages} {1825} (\bibinfo {year}
  {1985})}\BibitemShut {NoStop}%
\bibitem [{\citenamefont {Bernabei}\ \emph {et~al.}(1986)\citenamefont
  {Bernabei}, \citenamefont {Incicchitti}, \citenamefont {Mattioli},
  \citenamefont {Picozza}, \citenamefont {Prosperi}, \citenamefont {Casano},
  \citenamefont {d'Angelo}, \citenamefont {Pascale}, \citenamefont {Schaerf},
  \citenamefont {Giordano}, \citenamefont {Matone}, \citenamefont {Frullani},\
  and\ \citenamefont {Girolami}}]{Bern86}%
  \BibitemOpen
  \bibfield  {author} {\bibinfo {author} {\bibfnamefont {R.}~\bibnamefont
  {Bernabei}}, \bibinfo {author} {\bibfnamefont {A.}~\bibnamefont
  {Incicchitti}}, \bibinfo {author} {\bibfnamefont {M.}~\bibnamefont
  {Mattioli}}, \bibinfo {author} {\bibfnamefont {P.}~\bibnamefont {Picozza}},
  \bibinfo {author} {\bibfnamefont {D.}~\bibnamefont {Prosperi}}, \bibinfo
  {author} {\bibfnamefont {L.}~\bibnamefont {Casano}}, \bibinfo {author}
  {\bibfnamefont {S.}~\bibnamefont {d'Angelo}}, \bibinfo {author}
  {\bibfnamefont {M.~P.~D.}\ \bibnamefont {Pascale}}, \bibinfo {author}
  {\bibfnamefont {C.}~\bibnamefont {Schaerf}}, \bibinfo {author} {\bibfnamefont
  {G.}~\bibnamefont {Giordano}}, \bibinfo {author} {\bibfnamefont
  {G.}~\bibnamefont {Matone}}, \bibinfo {author} {\bibfnamefont
  {S.}~\bibnamefont {Frullani}}, \ and\ \bibinfo {author} {\bibfnamefont
  {B.}~\bibnamefont {Girolami}},\ }\href@noop {} {\bibfield  {journal}
  {\bibinfo  {journal} {Phys. Rev. Lett.}\ }\textbf {\bibinfo {volume} {57}},\
  \bibinfo {pages} {1542} (\bibinfo {year} {1986})}\BibitemShut {NoStop}%
\bibitem [{\citenamefont {Schiavilla}(2005)}]{Scia05}%
  \BibitemOpen
  \bibfield  {author} {\bibinfo {author} {\bibfnamefont {R.}~\bibnamefont
  {Schiavilla}},\ }\href@noop {} {\bibfield  {journal} {\bibinfo  {journal}
  {Phys. Rev. C}\ }\textbf {\bibinfo {volume} {72}},\ \bibinfo {pages} {034001}
  (\bibinfo {year} {2005})}\BibitemShut {NoStop}%
\bibitem [{\citenamefont {Arnold}\ \emph {et~al.}(2011)\citenamefont {Arnold},
  \citenamefont {Clegg}, \citenamefont {Karwowski}, \citenamefont {Rich},
  \citenamefont {Tompkins},\ and\ \citenamefont {Howell}}]{Arno11}%
  \BibitemOpen
  \bibfield  {author} {\bibinfo {author} {\bibfnamefont {C.~W.}\ \bibnamefont
  {Arnold}}, \bibinfo {author} {\bibfnamefont {T.~B.}\ \bibnamefont {Clegg}},
  \bibinfo {author} {\bibfnamefont {H.~J.}\ \bibnamefont {Karwowski}}, \bibinfo
  {author} {\bibfnamefont {G.~C.}\ \bibnamefont {Rich}}, \bibinfo {author}
  {\bibfnamefont {J.~R.}\ \bibnamefont {Tompkins}}, \ and\ \bibinfo {author}
  {\bibfnamefont {C.~R.}\ \bibnamefont {Howell}},\ }\href@noop {} {\bibfield
  {journal} {\bibinfo  {journal} {Nucl.\ Instrum.\ Methods Phys.\ Res. Sect.
  A}\ }\textbf {\bibinfo {volume} {647}},\ \bibinfo {pages} {55} (\bibinfo
  {year} {2011})}\BibitemShut {NoStop}%
\bibitem [{\citenamefont {Koning}\ and\ \citenamefont
  {Rochman}(2012)}]{Koni12}%
  \BibitemOpen
  \bibfield  {author} {\bibinfo {author} {\bibfnamefont {A.~J.}\ \bibnamefont
  {Koning}}\ and\ \bibinfo {author} {\bibfnamefont {D.}~\bibnamefont
  {Rochman}},\ }\href@noop {} {\bibfield  {journal} {\bibinfo  {journal} {Nucl.
  Data Sheets}\ }\textbf {\bibinfo {volume} {113}},\ \bibinfo {pages} {2841}
  (\bibinfo {year} {2012})}\BibitemShut {NoStop}%
\bibitem [{\citenamefont {P\'eru}\ and\ \citenamefont {Goutte}(2008)}]{Peru08}%
  \BibitemOpen
  \bibfield  {author} {\bibinfo {author} {\bibfnamefont {S.}~\bibnamefont
  {P\'eru}}\ and\ \bibinfo {author} {\bibfnamefont {H.}~\bibnamefont
  {Goutte}},\ }\href@noop {} {\bibfield  {journal} {\bibinfo  {journal} {Phys.
  Rev. C}\ }\textbf {\bibinfo {volume} {77}},\ \bibinfo {pages} {044313}
  (\bibinfo {year} {2008})}\BibitemShut {NoStop}%
\bibitem [{\citenamefont {Goriely}\ \emph {et~al.}(2008)\citenamefont
  {Goriely}, \citenamefont {Hilaire},\ and\ \citenamefont {Koning}}]{Gori08}%
  \BibitemOpen
  \bibfield  {author} {\bibinfo {author} {\bibfnamefont {S.}~\bibnamefont
  {Goriely}}, \bibinfo {author} {\bibfnamefont {S.}~\bibnamefont {Hilaire}}, \
  and\ \bibinfo {author} {\bibfnamefont {A.~J.}\ \bibnamefont {Koning}},\
  }\href@noop {} {\bibfield  {journal} {\bibinfo  {journal} {Phys. Rev. C}\
  }\textbf {\bibinfo {volume} {78}},\ \bibinfo {pages} {064307} (\bibinfo
  {year} {2008})}\BibitemShut {NoStop}%
\bibitem [{\citenamefont {Koning}\ and\ \citenamefont
  {Delaroche}(2003)}]{Koni03}%
  \BibitemOpen
  \bibfield  {author} {\bibinfo {author} {\bibfnamefont {A.~J.}\ \bibnamefont
  {Koning}}\ and\ \bibinfo {author} {\bibfnamefont {J.~P.}\ \bibnamefont
  {Delaroche}},\ }\href@noop {} {\bibfield  {journal} {\bibinfo  {journal}
  {Nucl.\ Phys.\ A}\ }\textbf {\bibinfo {volume} {713}},\ \bibinfo {pages}
  {231} (\bibinfo {year} {2003})}\BibitemShut {NoStop}%
\bibitem [{\citenamefont {Capote}\ \emph {et~al.}(2009)\citenamefont {Capote}
  \emph {et~al.}}]{Capo09}%
  \BibitemOpen
  \bibfield  {author} {\bibinfo {author} {\bibfnamefont {R.}~\bibnamefont
  {Capote}} \emph {et~al.},\ }\href@noop {} {\bibfield  {journal} {\bibinfo
  {journal} {Nucl. Data Sheets}\ }\textbf {\bibinfo {volume} {110}},\ \bibinfo
  {pages} {3107} (\bibinfo {year} {2009})}\BibitemShut {NoStop}%
\bibitem [{\citenamefont {Koning}\ \emph {et~al.}(2008)\citenamefont {Koning},
  \citenamefont {Hilaire},\ and\ \citenamefont {Goriely}}]{Koni08}%
  \BibitemOpen
  \bibfield  {author} {\bibinfo {author} {\bibfnamefont {A.~J.}\ \bibnamefont
  {Koning}}, \bibinfo {author} {\bibfnamefont {S.}~\bibnamefont {Hilaire}}, \
  and\ \bibinfo {author} {\bibfnamefont {S.}~\bibnamefont {Goriely}},\
  }\href@noop {} {\bibfield  {journal} {\bibinfo  {journal} {Nucl.\ Phys.\ A}\
  }\textbf {\bibinfo {volume} {810}},\ \bibinfo {pages} {13} (\bibinfo {year}
  {2008})}\BibitemShut {NoStop}%
\bibitem [{\citenamefont {Kopecky}\ and\ \citenamefont {Uhl}(1990)}]{Kope90}%
  \BibitemOpen
  \bibfield  {author} {\bibinfo {author} {\bibfnamefont {J.}~\bibnamefont
  {Kopecky}}\ and\ \bibinfo {author} {\bibfnamefont {M.}~\bibnamefont {Uhl}},\
  }\href@noop {} {\bibfield  {journal} {\bibinfo  {journal} {Phys. Rev. C}\
  }\textbf {\bibinfo {volume} {41}},\ \bibinfo {pages} {1941} (\bibinfo {year}
  {1990})}\BibitemShut {NoStop}%
\bibitem [{\citenamefont {Bauge}\ \emph {et~al.}(1998)\citenamefont {Bauge},
  \citenamefont {Delaroche},\ and\ \citenamefont {Girod}}]{Baug98}%
  \BibitemOpen
  \bibfield  {author} {\bibinfo {author} {\bibfnamefont {E.}~\bibnamefont
  {Bauge}}, \bibinfo {author} {\bibfnamefont {J.~P.}\ \bibnamefont
  {Delaroche}}, \ and\ \bibinfo {author} {\bibfnamefont {M.}~\bibnamefont
  {Girod}},\ }\href@noop {} {\bibfield  {journal} {\bibinfo  {journal} {Phys.
  Rev. C}\ }\textbf {\bibinfo {volume} {58}},\ \bibinfo {pages} {1118}
  (\bibinfo {year} {1998})}\BibitemShut {NoStop}%
\bibitem [{\citenamefont {Bauge}\ \emph {et~al.}(2001)\citenamefont {Bauge},
  \citenamefont {Delaroche},\ and\ \citenamefont {Girod}}]{Baug01}%
  \BibitemOpen
  \bibfield  {author} {\bibinfo {author} {\bibfnamefont {E.}~\bibnamefont
  {Bauge}}, \bibinfo {author} {\bibfnamefont {J.~P.}\ \bibnamefont
  {Delaroche}}, \ and\ \bibinfo {author} {\bibfnamefont {M.}~\bibnamefont
  {Girod}},\ }\href@noop {} {\bibfield  {journal} {\bibinfo  {journal} {Phys.
  Rev. C}\ }\textbf {\bibinfo {volume} {63}},\ \bibinfo {pages} {024607}
  (\bibinfo {year} {2001})}\BibitemShut {NoStop}%
\bibitem [{\citenamefont {Voght}\ \emph {et~al.}(2002)\citenamefont {Voght}
  \emph {et~al.}}]{Vogh02}%
  \BibitemOpen
  \bibfield  {author} {\bibinfo {author} {\bibfnamefont {K.}~\bibnamefont
  {Voght}} \emph {et~al.},\ }\href@noop {} {\bibfield  {journal} {\bibinfo
  {journal} {Nucl.\ Phys.\ A}\ }\textbf {\bibinfo {volume} {707}},\ \bibinfo
  {pages} {241} (\bibinfo {year} {2002})}\BibitemShut {NoStop}%
\bibitem [{\citenamefont {Beil}\ \emph {et~al.}(1974)\citenamefont {Beil},
  \citenamefont {Berg\'ere}, \citenamefont {Carlos}, \citenamefont
  {Lepr\^etre}, \citenamefont {de~Miniac},\ and\ \citenamefont
  {Veyssi\'ere}}]{Beil74}%
  \BibitemOpen
  \bibfield  {author} {\bibinfo {author} {\bibfnamefont {H.}~\bibnamefont
  {Beil}}, \bibinfo {author} {\bibfnamefont {R.}~\bibnamefont {Berg\'ere}},
  \bibinfo {author} {\bibfnamefont {P.}~\bibnamefont {Carlos}}, \bibinfo
  {author} {\bibfnamefont {A.}~\bibnamefont {Lepr\^etre}}, \bibinfo {author}
  {\bibfnamefont {A.}~\bibnamefont {de~Miniac}}, \ and\ \bibinfo {author}
  {\bibfnamefont {A.}~\bibnamefont {Veyssi\'ere}},\ }\href@noop {} {\bibfield
  {journal} {\bibinfo  {journal} {Nucl.\ Phys.\ A}\ }\textbf {\bibinfo {volume}
  {227}},\ \bibinfo {pages} {427} (\bibinfo {year} {1974})}\BibitemShut
  {NoStop}%
\bibitem [{\citenamefont {Berman}\ \emph {et~al.}(1967)\citenamefont {Berman},
  \citenamefont {Caldwell}, \citenamefont {Harvey}, \citenamefont {Kelly},
  \citenamefont {Bramblett},\ and\ \citenamefont {Fultz}}]{Berm67}%
  \BibitemOpen
  \bibfield  {author} {\bibinfo {author} {\bibfnamefont {B.~L.}\ \bibnamefont
  {Berman}}, \bibinfo {author} {\bibfnamefont {J.~T.}\ \bibnamefont
  {Caldwell}}, \bibinfo {author} {\bibfnamefont {R.~R.}\ \bibnamefont
  {Harvey}}, \bibinfo {author} {\bibfnamefont {M.~A.}\ \bibnamefont {Kelly}},
  \bibinfo {author} {\bibfnamefont {R.~L.}\ \bibnamefont {Bramblett}}, \ and\
  \bibinfo {author} {\bibfnamefont {S.~C.}\ \bibnamefont {Fultz}},\ }\href@noop
  {} {\bibfield  {journal} {\bibinfo  {journal} {Phys. Rev.}\ }\textbf
  {\bibinfo {volume} {162}},\ \bibinfo {pages} {1098} (\bibinfo {year}
  {1967})}\BibitemShut {NoStop}%
\bibitem [{\citenamefont {Lepr\^etre}\ \emph {et~al.}(1971)\citenamefont
  {Lepr\^etre}, \citenamefont {Beil}, \citenamefont {Berg\'ere}, \citenamefont
  {Carlos},\ and\ \citenamefont {Veyssi\'ere}}]{Lepr71}%
  \BibitemOpen
  \bibfield  {author} {\bibinfo {author} {\bibfnamefont {A.}~\bibnamefont
  {Lepr\^etre}}, \bibinfo {author} {\bibfnamefont {H.}~\bibnamefont {Beil}},
  \bibinfo {author} {\bibfnamefont {R.}~\bibnamefont {Berg\'ere}}, \bibinfo
  {author} {\bibfnamefont {P.}~\bibnamefont {Carlos}}, \ and\ \bibinfo {author}
  {\bibfnamefont {A.}~\bibnamefont {Veyssi\'ere}},\ }\href@noop {} {\bibfield
  {journal} {\bibinfo  {journal} {Nucl.\ Phys.\ A}\ }\textbf {\bibinfo {volume}
  {175}},\ \bibinfo {pages} {609} (\bibinfo {year} {1971})}\BibitemShut
  {NoStop}%
\bibitem [{\citenamefont {Berman}\ \emph {et~al.}(1987)\citenamefont {Berman},
  \citenamefont {Pywell}, \citenamefont {Dietrich}, \citenamefont {Thompson},
  \citenamefont {McNeill},\ and\ \citenamefont {Jury}}]{Berm87}%
  \BibitemOpen
  \bibfield  {author} {\bibinfo {author} {\bibfnamefont {B.~L.}\ \bibnamefont
  {Berman}}, \bibinfo {author} {\bibfnamefont {R.~E.}\ \bibnamefont {Pywell}},
  \bibinfo {author} {\bibfnamefont {S.~S.}\ \bibnamefont {Dietrich}}, \bibinfo
  {author} {\bibfnamefont {M.~N.}\ \bibnamefont {Thompson}}, \bibinfo {author}
  {\bibfnamefont {K.~G.}\ \bibnamefont {McNeill}}, \ and\ \bibinfo {author}
  {\bibfnamefont {J.~W.}\ \bibnamefont {Jury}},\ }\href@noop {} {\bibfield
  {journal} {\bibinfo  {journal} {Phys. Rev. C}\ }\textbf {\bibinfo {volume}
  {36}},\ \bibinfo {pages} {1286} (\bibinfo {year} {1987})}\BibitemShut
  {NoStop}%
\bibitem [{\citenamefont {Varlamov}\ \emph {et~al.}(2009)\citenamefont
  {Varlamov}, \citenamefont {Peskov},\ and\ \citenamefont {Stepanov}}]{Varl09}%
  \BibitemOpen
  \bibfield  {author} {\bibinfo {author} {\bibfnamefont {V.~V.}\ \bibnamefont
  {Varlamov}}, \bibinfo {author} {\bibfnamefont {N.~N.}\ \bibnamefont
  {Peskov}}, \ and\ \bibinfo {author} {\bibfnamefont {M.~E.}\ \bibnamefont
  {Stepanov}},\ }\href@noop {} {\bibfield  {journal} {\bibinfo  {journal}
  {Phys. of At. Nuclei}\ }\textbf {\bibinfo {volume} {72}},\ \bibinfo {pages}
  {214} (\bibinfo {year} {2009})}\BibitemShut {NoStop}%
\bibitem [{Exf()}]{Exfo}%
  \BibitemOpen
  \href@noop {} {\enquote {\bibinfo {title} {http://www-nds.iaea.org/exfor},}\
  }\BibitemShut {NoStop}%
\bibitem [{\citenamefont {Schwengner}\ \emph {et~al.}(2008)\citenamefont
  {Schwengner}, \citenamefont {Rusev}, \citenamefont {Tsoneva}, \citenamefont
  {Benouaret}, \citenamefont {Beyer}, \citenamefont {Erhard}, \citenamefont
  {Grosse}, \citenamefont {Junghans}, \citenamefont {Klug}, \citenamefont
  {Kosev}, \citenamefont {Lenske}, \citenamefont {Nair}, \citenamefont
  {Schilling},\ and\ \citenamefont {Wagner}}]{Schw08}%
  \BibitemOpen
  \bibfield  {author} {\bibinfo {author} {\bibfnamefont {R.}~\bibnamefont
  {Schwengner}}, \bibinfo {author} {\bibfnamefont {G.}~\bibnamefont {Rusev}},
  \bibinfo {author} {\bibfnamefont {N.}~\bibnamefont {Tsoneva}}, \bibinfo
  {author} {\bibfnamefont {N.}~\bibnamefont {Benouaret}}, \bibinfo {author}
  {\bibfnamefont {R.}~\bibnamefont {Beyer}}, \bibinfo {author} {\bibfnamefont
  {M.}~\bibnamefont {Erhard}}, \bibinfo {author} {\bibfnamefont
  {E.}~\bibnamefont {Grosse}}, \bibinfo {author} {\bibfnamefont {A.~R.}\
  \bibnamefont {Junghans}}, \bibinfo {author} {\bibfnamefont {J.}~\bibnamefont
  {Klug}}, \bibinfo {author} {\bibfnamefont {K.}~\bibnamefont {Kosev}},
  \bibinfo {author} {\bibfnamefont {H.}~\bibnamefont {Lenske}}, \bibinfo
  {author} {\bibfnamefont {C.}~\bibnamefont {Nair}}, \bibinfo {author}
  {\bibfnamefont {K.~D.}\ \bibnamefont {Schilling}}, \ and\ \bibinfo {author}
  {\bibfnamefont {A.}~\bibnamefont {Wagner}},\ }\href@noop {} {\bibfield
  {journal} {\bibinfo  {journal} {Phys. Rev. C}\ }\textbf {\bibinfo {volume}
  {78}},\ \bibinfo {pages} {064314} (\bibinfo {year} {2008})}\BibitemShut
  {NoStop}%
\bibitem [{\citenamefont {Rusev}\ \emph {et~al.}(2009)\citenamefont {Rusev},
  \citenamefont {Schwengner}, \citenamefont {Beyer}, \citenamefont {Erhard},
  \citenamefont {Grosse}, \citenamefont {Junghans}, \citenamefont {Kosev},
  \citenamefont {Nair}, \citenamefont {Schilling}, \citenamefont {Wagner},
  \citenamefont {D\"onau},\ and\ \citenamefont {Frauendorf}}]{Ruse09}%
  \BibitemOpen
  \bibfield  {author} {\bibinfo {author} {\bibfnamefont {G.}~\bibnamefont
  {Rusev}}, \bibinfo {author} {\bibfnamefont {R.}~\bibnamefont {Schwengner}},
  \bibinfo {author} {\bibfnamefont {R.}~\bibnamefont {Beyer}}, \bibinfo
  {author} {\bibfnamefont {M.}~\bibnamefont {Erhard}}, \bibinfo {author}
  {\bibfnamefont {E.}~\bibnamefont {Grosse}}, \bibinfo {author} {\bibfnamefont
  {A.~R.}\ \bibnamefont {Junghans}}, \bibinfo {author} {\bibfnamefont
  {K.}~\bibnamefont {Kosev}}, \bibinfo {author} {\bibfnamefont
  {C.}~\bibnamefont {Nair}}, \bibinfo {author} {\bibfnamefont {K.~D.}\
  \bibnamefont {Schilling}}, \bibinfo {author} {\bibfnamefont {A.}~\bibnamefont
  {Wagner}}, \bibinfo {author} {\bibfnamefont {F.}~\bibnamefont {D\"onau}}, \
  and\ \bibinfo {author} {\bibfnamefont {S.}~\bibnamefont {Frauendorf}},\
  }\href@noop {} {\bibfield  {journal} {\bibinfo  {journal} {Phys. Rev. C}\
  }\textbf {\bibinfo {volume} {79}},\ \bibinfo {pages} {061302} (\bibinfo
  {year} {2009})}\BibitemShut {NoStop}%
\bibitem [{\citenamefont {Guttormsen}\ \emph {et~al.}(2005)\citenamefont
  {Guttormsen}, \citenamefont {Chankova}, \citenamefont {Agvaanluvsan},
  \citenamefont {Algin}, \citenamefont {Bernstein}, \citenamefont
  {Ingebretsen}, \citenamefont {L\"onnroth}, \citenamefont {Messelt},
  \citenamefont {Mitchell}, \citenamefont {Rekstad}, \citenamefont {Schiller},
  \citenamefont {Siem}, \citenamefont {Sunde}, \citenamefont {Voinov},\ and\
  \citenamefont {{\O}deg{\aa}rd}}]{Gutt05}%
  \BibitemOpen
  \bibfield  {author} {\bibinfo {author} {\bibfnamefont {M.}~\bibnamefont
  {Guttormsen}}, \bibinfo {author} {\bibfnamefont {R.}~\bibnamefont
  {Chankova}}, \bibinfo {author} {\bibfnamefont {U.}~\bibnamefont
  {Agvaanluvsan}}, \bibinfo {author} {\bibfnamefont {E.}~\bibnamefont {Algin}},
  \bibinfo {author} {\bibfnamefont {L.~A.}\ \bibnamefont {Bernstein}}, \bibinfo
  {author} {\bibfnamefont {F.}~\bibnamefont {Ingebretsen}}, \bibinfo {author}
  {\bibfnamefont {T.}~\bibnamefont {L\"onnroth}}, \bibinfo {author}
  {\bibfnamefont {S.}~\bibnamefont {Messelt}}, \bibinfo {author} {\bibfnamefont
  {G.~E.}\ \bibnamefont {Mitchell}}, \bibinfo {author} {\bibfnamefont
  {J.}~\bibnamefont {Rekstad}}, \bibinfo {author} {\bibfnamefont
  {A.}~\bibnamefont {Schiller}}, \bibinfo {author} {\bibfnamefont
  {S.}~\bibnamefont {Siem}}, \bibinfo {author} {\bibfnamefont {A.~C.}\
  \bibnamefont {Sunde}}, \bibinfo {author} {\bibfnamefont {A.}~\bibnamefont
  {Voinov}}, \ and\ \bibinfo {author} {\bibnamefont {{\O}deg{\aa}rd}},\
  }\href@noop {} {\bibfield  {journal} {\bibinfo  {journal} {Phys. Rev. C}\
  }\textbf {\bibinfo {volume} {71}},\ \bibinfo {pages} {044307} (\bibinfo
  {year} {2005})}\BibitemShut {NoStop}%
\bibitem [{\citenamefont {Rauscher}(2011)}]{Raus11}%
  \BibitemOpen
  \bibfield  {author} {\bibinfo {author} {\bibfnamefont {T.}~\bibnamefont
  {Rauscher}},\ }\href@noop {} {\bibfield  {journal} {\bibinfo  {journal}
  {Astroph. J.}\ }\textbf {\bibinfo {volume} {738}},\ \bibinfo {pages} {143}
  (\bibinfo {year} {2011})}\BibitemShut {NoStop}%
\end{thebibliography}%

\end{document}